\documentclass[aip,reprint,jcp]{revtex4-2}

\usepackage[version=3]{mhchem} % Formula subscripts using \ce{}
\usepackage[%
  colorlinks=true,
  urlcolor=blue,
  linkcolor=blue,
  citecolor=black
]{hyperref}
\usepackage{graphicx}
\usepackage{dcolumn}
\usepackage[normalem]{ulem}
\usepackage{bm}
\usepackage[utf8]{inputenc}
\usepackage[T1]{fontenc}
\usepackage{mathptmx}
\usepackage{amsmath}
\renewcommand{\selectlanguage}[1]{}
\DeclareUnicodeCharacter{2009}{\,}
\DeclareUnicodeCharacter{2032}{'}

\usepackage{xcolor}

\makeatletter
\def\@email#1#2{%
 \endgroup
 \patchcmd{\titleblock@produce}
  {\frontmatter@RRAPformat}
  {\frontmatter@RRAPformat{\produce@RRAP{*#1\href{mailto:#2}{#2}}}\frontmatter@RRAPformat}
  {}{}
}%
\makeatother
\begin{document}

\author{Ananya Venkatachalam}
\altaffiliation{These authors contributed equally.}
\affiliation{Department of Chemistry, Harvey Mudd College, Claremont, CA 91711, United States}

\author{Pierre J. Walker}
\altaffiliation{These authors contributed equally.}
\affiliation{Division of Chemistry and Chemical Engineering, California Institute of Technology, Pasadena, CA 91125, United States}

\author{Samuel Varner}
\altaffiliation{These authors contributed equally.}
\affiliation{Department of Chemical and Biomolecular Engineering, Tandon School of Engineering, New York University, Brooklyn, NY 11201, United States}

\author{Bilin Zhuang}
\affiliation{Department of Chemistry, Harvey Mudd College, Claremont, CA 91711, United States}
\email{bzhuang@hmc.edu}

\author{Zhen-Gang Wang}
\affiliation{Division of Chemistry and Chemical Engineering, California Institute of Technology, Pasadena, CA 91125, United States}
\email{zgw@caltech.edu}

%%%%%%%%%%%%%%%%%%%%%%%%%%%%%%%%%%%%%%%%%%%%%%%%%%%%%%%%%%%%%%%%%%%%%
%% The document title should be given as usual. Some journals require
%% a running title from the author: this should be supplied as an
%% optional argument to \title.
%%%%%%%%%%%%%%%%%%%%%%%%%%%%%%%%%%%%%%%%%%%%%%%%%%%%%%%%%%%%%%%%%%%%%
% \title[Dipole asymmetry and equilibrium properties]
%   {Effects of Dipole Positional Asymmetry: Equilibrium properties of a shifted-dipole Stockmayer fluid}
\title{Stockmayer Fluid with a Shifted Dipole. III. Ion Solvation}

\begin{abstract}

% with shift:
% long range --> change in solvation environment leads to less screening
% short range --> change in solvation environment leads to more
% "shielding"
% prioritize talking about solvent reorganization

% solvent around ion!!!

Though ion solvation is well understood for the standard Stockmayer fluid, the effect of displacing the point dipole of a Stockmayer particle, a simple geometric modification, on ion--solvent structuring and solvation thermodynamics remains less understood. We use molecular dynamics simulations to describe how a dipole shift affects radial and orientational solvent structure around centered ions, how strongly the solvent distinguishes cations and anions, and how the solvent mediates ion--ion pairing. We show that, with a centered dipole, cations and anions have equal preferences for the local solvent environment. Shifting the dipole breaks the fore--aft symmetry of the local electrostatic field, leading to a stronger preference for a Stockmayer particle's dipole to orient towards an anion and a weaker preference to orient away from a cation. These structural asymmetries propagate to the solvation free energies: increasing dipole displacement makes anion solvation more favorable, while cation solvation becomes less favorable overall across dipole strengths. The same asymmetry also manifests in the effective ion-ion interaction---increasing the dipole shift reorganizes the local solvation environment, enhancing solvent shielding of short-range cation--anion contact while weakening screening at longer range. As a result, solvent-separated configurations become comparatively more favorable, whereas increasing dipole strength produces the opposite trend. Such effects cannot be explained by bulk dielectric screening alone, indicating that molecular charge geometry plays an important role in both single-ion solvation and solvent-mediated ion--ion interactions. This work demonstrates that dipole location, not only magnitude, provides a geometric control on ion-specific solvation and ion pairing in general polar fluids.

\end{abstract}

\maketitle

\section{Introduction}

% Electrolyte solutions are a key example of how molecular electrostatics propagate local liquid structure to larger-scale thermodynamic behavior. 

Molecular electrostatics organize local liquid structure and connect molecular-scale interactions to larger-scale thermodynamic behavior. This connection is especially important in determining the local solvation environments within electrolyte solutions. For example, around an ion, the arrangement of nearby solvent molecules determines hydration structure\cite{shockMolecularDynamicsSimulations2023,briniHowWatersProperties2017,fongInterplaySolvationPolarization2024}, solvation free energy\cite{shockSolvationEnergyIons2020,basdoganMachineLearningGuidedApproach2020}, and effective interactions with other charged species\cite{briniHowWatersProperties2017,varnerEntropicOriginIonic2023,oneillPairNotPair2024}. At interfaces, the same local solvation physics can become even more important because of how ion concentrations, solvent orientation, and electrostatic field can vary across the interfacial region~\cite{nakamuraIonSolvationLiquid2012,gregoryUnderstandingSpecificIon2022,jungwirthSpecificIonEffects2006, tobiasSimulationTheoryIons2013, bastos-gonzalezIonsInterfacesCentral2016,simonsObservationsElectronicCharacter2024}. Such behavior is relevant to electrolyte-containing systems ranging from atmospheric aerosols\cite{nakamuraIonSolvationLiquid2012,tobiasSimulationTheoryIons2013}, ocean surfaces\cite{bjorneholmWaterInterfaces2016,carpenterChemistryReleaseGases2015}, reactions at interfaces\cite{benjaminChemicalReactionsSolvation1996,xiongStrongElectricField2020,liuStrongElectricField2024,shiWaterStructureElectric2025,beckerElectrokineticElectrochemicalElectrostatic2022,zhangDoubleLayerDistributionHydronium2024} and battery electrolytes\cite{fongSolvationShellsPower2025,oneillPairNotPair2024,houSolvationStructureTransport2021,jonesInferringGlobalDynamics2022,zengModulatingSolvationStructure2025,chengEmergingEraElectrolyte2022,tianElectrolyteSolvationStructure2022,murphyInterfacialStructureIon2026}. Despite the importance of ions in these cases, it remains difficult to isolate the role of solvent dipolar asymmetry, which we have already shown is of incredible importance to bulk~\cite{walkerStockmayerFluidShifted2026a} and interfacial~\cite{varnerStockmayerFluidShifted2026} behavior of the pure solvent, from the many other molecular features that influence ion solvation.

A demonstration of this knowledge gap is the long-standing phenomenon of the ionic ``sign effect"\cite{kusakaIoninducedNucleationDensity1995,hantalDoesSignCharge2023}, in which cations and anions can display different hydration and nucleation behavior even when their sizes and non-electrostatic interactions are otherwise similar or identical. Early studies of condensation on gaseous ions has already suggested that positive and negative ions could promote droplet formation differently, implying that the surrounding polar medium does not necessarily respond symmetrically to opposite charges~\cite{wilsonXICondensationWater1897, loebNatureIonicSign1938, russellNucleationGaseousIons1969}. More recent simulations and theories have extended this question to aqueous interfaces and electrolyte solutions, where ion charge sign can impact surface affinity and thus reaction dynamics~\cite{jungwirthSpecificIonEffects2006,tobiasSimulationTheoryIons2013,hantalDoesSignCharge2023}. Atomistic water simulations provide insight into such effects, showing, for example, that two ions differing only in charge sign can have distinct partitioning tendencies at interfaces~\cite{hantalDoesSignCharge2023}. For sufficiently small ions, anions may be more significantly hydrated because hydrogens on water can approach a negative ion more closely than the oxygen atom can approach a positive ion; for larger ions, this trend can reverse~\cite{hantalDoesSignCharge2023}. Related sign-dependent behavior has also been observed in polyelectrolyte solutions, where otherwise comparable anionic and cationic systems can form different local structures~\cite{ohNucleationWaterMethanol2001,linStructuralPropertiesIons2024}. The described examples suggest that, rather than being a property of a given ion alone, the sign effect reflects how the solvent reorganizes around positive and negative charges, thus raising the question: how much can this asymmetric response be explained by the geometry of the solvent charge distribution itself?

Minimal dipolar fluids offer a controlled setting to address this question. The ordinary Stockmayer fluid, consisting of Lennard-Jones particles with permanent point dipoles, has previously been used as a simple model for polar liquids~\cite{stockmayerSecondVirialCoefficients1941,shockSolvationEnergyIons2020,shockMolecularDynamicsSimulations2023,liuMinimalCoarseGrainedModels2024}. This model allows for the examination of pair correlations, dielectric response, and solvation in an environment where electrostatic and excluded-volume effects can be interpreted directly, with no chemical specificity~\cite{shockSolvationEnergyIons2020,shockMolecularDynamicsSimulations2023}. However, the standard Stockmayer fluid places the dipole at the center of a spherical particle, resulting in fore--aft symmetry and therefore identical solvation environments for equal-sized cations and anions, apart from the reversal of dipole orientation. As a result, the ordinary Stockmayer fluid is insufficient to test whether solvent dipolar asymmetry alone can generate sign-specific ion solvation.

By displacing the dipole from the Lennard-Jones center, the \textit{shifted-dipole} Stockmayer fluid distinguishes the two sides of the solvent molecule, introducing a controlled dipolar asymmetry while retaining a minimal electrostatic description. This model was first introduced by Kusaka, Wang, and Seinfeld to study sign effects in ion-induced nucleation~\cite{kusakaIoninducedNucleationDensity1995}. The liquid--vapor phase behavior and dielectric response of the shifted-dipole Stockmayer fluid have been examined using the Co-Oriented Fluid Functional Equation for Electrostatic Interactions (COFFEE), which explicitly incorporates molecular orientational structure~\cite{langenbachCoOrientedFluidFunctional2017,langenbachRelativePermittivityDipolar2020}. More recently, studies of the bulk and liquid--vapor interfacial behavior of this model established that even modest dipole shifts can reorganize angular correlations, alter dielectric response, and generate interfacial polarization~\cite{walkerStockmayerFluidShifted2026a,varnerStockmayerFluidShifted2026,fertigBasicRequirementsPotential2025}. Related shifted-dipole models have also been used to study self-assembly and clustering in ferrofluids and anisotropic colloidal systems~\cite{kantorovichFerrofluidsShiftedDipoles2011,kantorovichMagneticParticlesShifted2011,klinkigtClusterFormationSystems2013,weeberMicrostructureMagneticProperties2013}. These results motivate the next question: how does molecular dipolar asymmetry affect the solvation of explicit ions?

In this work, we use molecular dynamics simulations to examine and compare the solvation of cations and anions in the shifted-dipole Stockmayer fluid. We first analyze ion--solvent radial distribution functions to determine how the solvation shell structure, as well as the coordination number, changes with varying dipole strength and shift. We then use angular--radial distribution functions to determine how solvent dipoles orient around each ion. To connect this structural information to thermodynamics, we compute the change in solvation free energy and compare the resulting cation--anion asymmetry across solvent parameters. Finally, we examine how the shifted dipole solvent mediates ion interactions by calculating the potential of mean force between oppositely charged ions. As a whole, the observables allow us to determine how dipolar asymmetry alone affects the observed ionic sign effect and resulting solvation thermodynamics.

\begin{figure}[ht!]
    \centering
        \includegraphics[width=0.5\linewidth]{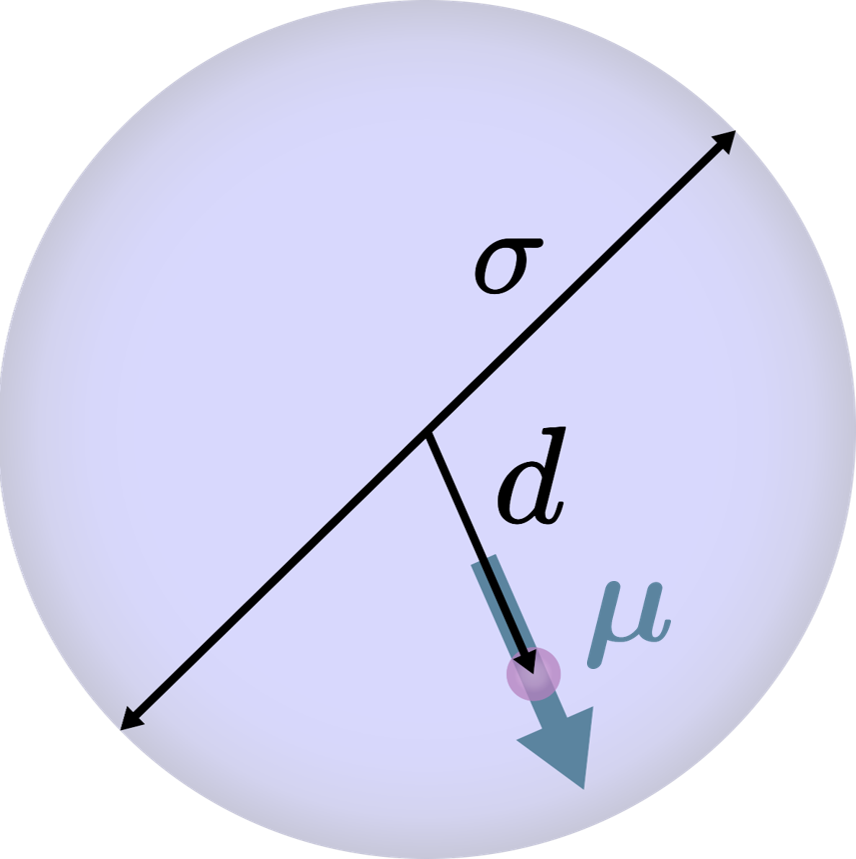}    
    \caption{Model of a shifted-dipole Stockmayer fluid (sd-SF). Compared to the regular Stockmayer fluid, the particle has a dipole $\bm{\mu}$ shifted from the center by a distance $d$ in the direction of the dipole (depicted as the blue arrow). The excluded-volume interaction between the particles is described by a regular Lennard-Jones potential with characteristic length scale $\sigma$, and the electrostatic interaction is the dipole-dipole interaction between the shifted dipoles. Figure adapted and caption reproduced from Walker et al.~\cite{walkerStockmayerFluidShifted2026a}.}  
    \label{fig:shifted-stockmayer}
\end{figure}

\section{Methodology}

\subsection{System}

We study a system of shifted-dipole Stockmayer (sd-SF) solvent particles with particle number $N$, at fixed box volume $V$, and temperature $T$. Each solvent particle $i$, illustrated schematically in Fig.~\ref{fig:shifted-stockmayer}, has a Lennard-Jones (LJ) center at position $\mathbf{r}_i$ and a permanent dipole moment $\pmb{\mu}_i$ displaced from this center by a distance $d$ along the dipole vector. The position of the displaced dipole is thus $\mathbf{r}^{d}_i = \mathbf{r}_i +  \hat{\pmb{\mu}}_i d$, where $\hat{\pmb{\mu}}_i$ is a unit vector parallel to $\pmb{\mu}_i$. The solvent--solvent non-electrostatic interaction is a LJ potential between molecular centers:
\begin{equation}
    \label{eq:lj}
    \phi_{\text{LJ}}(r_{ij}) = 
    4\epsilon \left[ \left( \frac{\sigma}{r_{ij}} \right)^{12} - \left( \frac{\sigma}{r_{ij}} \right)^6 \right]\,,
\end{equation}
where $r_{ij} = |\mathbf{r}_j - \mathbf{r}_i|$ is the distance between the geometrical centers of particles $i$ and $j$, $\epsilon$ is the pair interaction strength that sets the energy scale, and $\sigma$ is the characteristic particle diameter. 
% We set the LJ cut-off distance to $r_c = 2.5\sigma$, which is well beyond the potential minimum at $r_{\text{min}} = 2^{1/6}\sigma \approx 1.122 \sigma$. 

The permanent point dipoles interact through the dipole--dipole potential evaluated between the displaced dipole sites:
\begin{equation}
\label{dip}
\phi_{\mathrm{dipole}}(\mathbf r_{ij}^d)
=
\frac{1}{4\pi\epsilon_0}
\left[
\frac{\pmb{\mu_i^*}\cdot\pmb{\mu_j^*}}{(r_{ij}^d)^3}
-
3
\frac{
\left(\pmb{\mu_i^*}\cdot\mathbf r_{ij}^d\right)
\left(\pmb{\mu_j^*}\cdot\mathbf r_{ij}^d\right)
}{(r_{ij}^d)^5}
\right],
\end{equation}
where $\mathbf r_{ij}^d=\mathbf r_j^d-\mathbf r_i^d$ and $r_{ij}^d=|\mathbf r_{ij}^d|$. When $d=0$, this model reduces to the ordinary Stockmayer fluid, while finite $d$ breaks the fore--aft symmetry of the solvent molecule by separating the Lennard-Jones center from the electrostatic center\cite{stockmayerSecondVirialCoefficients1941}.
 
Ions are represented as spherical particles of diameter $\sigma$ with point charges located at their centers. Using LJ reduced units, we define the charge magnitude as $q^* =
\frac{q}{\sqrt{4\pi\epsilon_0\epsilon\sigma}}$. The cations and anions considered here have equal size and equal charge magnitude with opposite sign.
This choice allows us to isolate charge-sign effects arising from solvent asymmetry rather than from differences in ion size or short-range ion--solvent interactions.

% Compared with the bulk solvent simulations described in our previous paper \cite{walkerStockmayerFluidShifted2026}, the ionic simulations required a few additional considerations. Because explicit ions introduce charge--charge and charge--dipole interactions in addition to the original dipole--dipole forces, the long-range electrostatic treatment was modified so that all electrostatic interactions could be handled consistently. Introducing ions also adds new short-range excluded-volume interactions, both between ions and solvent molecules and between pairs of ions. The ion--ion Lennard-Jones interactions were truncated at $r_c=2^{1/6}\sigma$ to remain purely repulsive and represent only excluded-volume effects, a choice that isolates the role of electrostatics by preventing the ions from experiencing an additional short-range attractive interaction. As in the bulk simulations, the solvent molecules were treated as rigid two-site particles, while the ions were propagated separately as single spherical particles. Because the solvent molecule is charge-asymmetric, cations and anions experience fundamentally different ion--dipole interactions, making sign asymmetry the primary focus of the ionic simulations.

\subsection{Molecular Dynamics}
\label{sec:md}
\subsubsection{Simulations}

We perform molecular dynamics (MD) simulations of the shifted-dipole Stockmayer model in the NVT ensemble using the simulation package LAMMPS~\cite {thompsonLAMMPSFlexibleSimulation2022}, executed on graphics processing units (GPUs)~\cite{brownImplementingMolecularDynamics2011,brownImplementingMolecularDynamics2012}. Because no shifted-dipole Stockmayer potential is readily available in LAMMPS, we represent each solvent particle $i$ as a rigid dimer composed of two components (Fig.~\ref{fig:shifted-stockmayer}): an LJ particle with characteristic size $\sigma$ and a massless, infinitesimally small particle carrying permanent dipole moment $\boldsymbol{\mu}_i$. The sites are separated by a fixed bond length $d$, so that the dipole is displaced from the geometric center of the particle. The massless particle's dipole is constrained to remain parallel to the rigid bond, pointing radially outward, and interacts only through dipolar forces. In the limit of no shift ($d$=0), this model reduces to the standard Stockmayer fluid, while remaining suitable for physically relevant ranges of dipole displacement and magnitude. To elucidate the role of dipole magnitude and shift on the local solvation environment of cations and anions, we systematically vary $\mu = |\boldsymbol{\mu}|$ and $d$, respectively.

Note that, to avoid a numerical discontinuity at the LJ cutoff, we apply a constant shift $S$ to the potential:
\begin{equation}
    \label{eq:lj_s}
    \phi_{\text{LJ}}^{\text{shifted}}(r_{ij}) = 
    \begin{cases} 
    4\epsilon \left[ \left( \frac{\sigma}{r_{ij}} \right)^{12} - \left( \frac{\sigma}{r_{ij}} \right)^6 \right] + S, & \text{if } r_{ij} \leq r_c\,,\\
    0, & \text{if } r_{ij} > r_c \,,
    \end{cases}
\end{equation}
where $S$ is chosen such that $\phi_{\text{LJ}}^{\text{shifted}}(r_c)=0$. As in our prior work~\cite{varnerStockmayerFluidShifted2026,walkerStockmayerFluidShifted2026a}, dipolar interactions are evaluated in real space up to a cutoff distance of $8.0\sigma$, beyond which they are computed in Fourier space using the particle--particle particle--mesh (PPPM) method for long--range dipolar interactions \cite{cerdaP3MAlgorithmDipolar2008}.

The system densities used in the present work are obtained from the saturated liquid phase of the pure solvents determined from the liquid--vapor slab simulations described in our previous studies of the shifted-dipole Stockmayer fluid~\cite{varnerStockmayerFluidShifted2026,walkerStockmayerFluidShifted2026a}. The full procedure used to form and equilibrate the liquid slab, apply the slab correction, and determine the coexistence liquid density for each value of $\mu$ and $d$ is described in Refs.~\onlinecite{varnerStockmayerFluidShifted2026,walkerStockmayerFluidShifted2026a} and provided in the supporting information (SI). The resulting liquid densities are used here to set the dimensions of the bulk simulation boxes.

We carried out bulk simulations at the liquid coexistence density, with the number of ions and solvent particles varied based on the calculation. Depending on the observable, systems contained either a single ion, one cation--anion pair, or multiple ion pairs. The simulations employ a Nos\'e--Hoover--type thermostat that accounts for both translational and rotational degrees of freedom in rigid bodies~\cite{kamberajTimeReversibleSymplectic2005,hooverCanonicalDynamicsEquilibrium1985,martynaExplicitReversibleIntegrators1996,martynaNoseHooverChains1992}, with a damping parameter $\tau = 100\Delta t$ and a time step of $\Delta t = 0.005\tau_0$, where $\tau_0=\sqrt{m\sigma^2/\epsilon}$.

\subsubsection{Structural Properties}

As in our previous work \cite{walkerStockmayerFluidShifted2026a}, we analyze simulation trajectories using MDAnalysis \cite{michaud-agrawalMDAnalysisToolkitAnalysis2011} and MDCraft \cite{yeMDCraftPythonAssistant2024}, the latter of which can accommodate the features of the shifted-dipole Stockmayer model through a custom trajectory reader. In this work, which involves a dilute concentration of ions in a solution of sd-SF particles, we focus on positional and orientational correlations between ions and sd-SF particles. 

\begin{figure}[ht]
    \centering
    \includegraphics[width=0.7\linewidth]{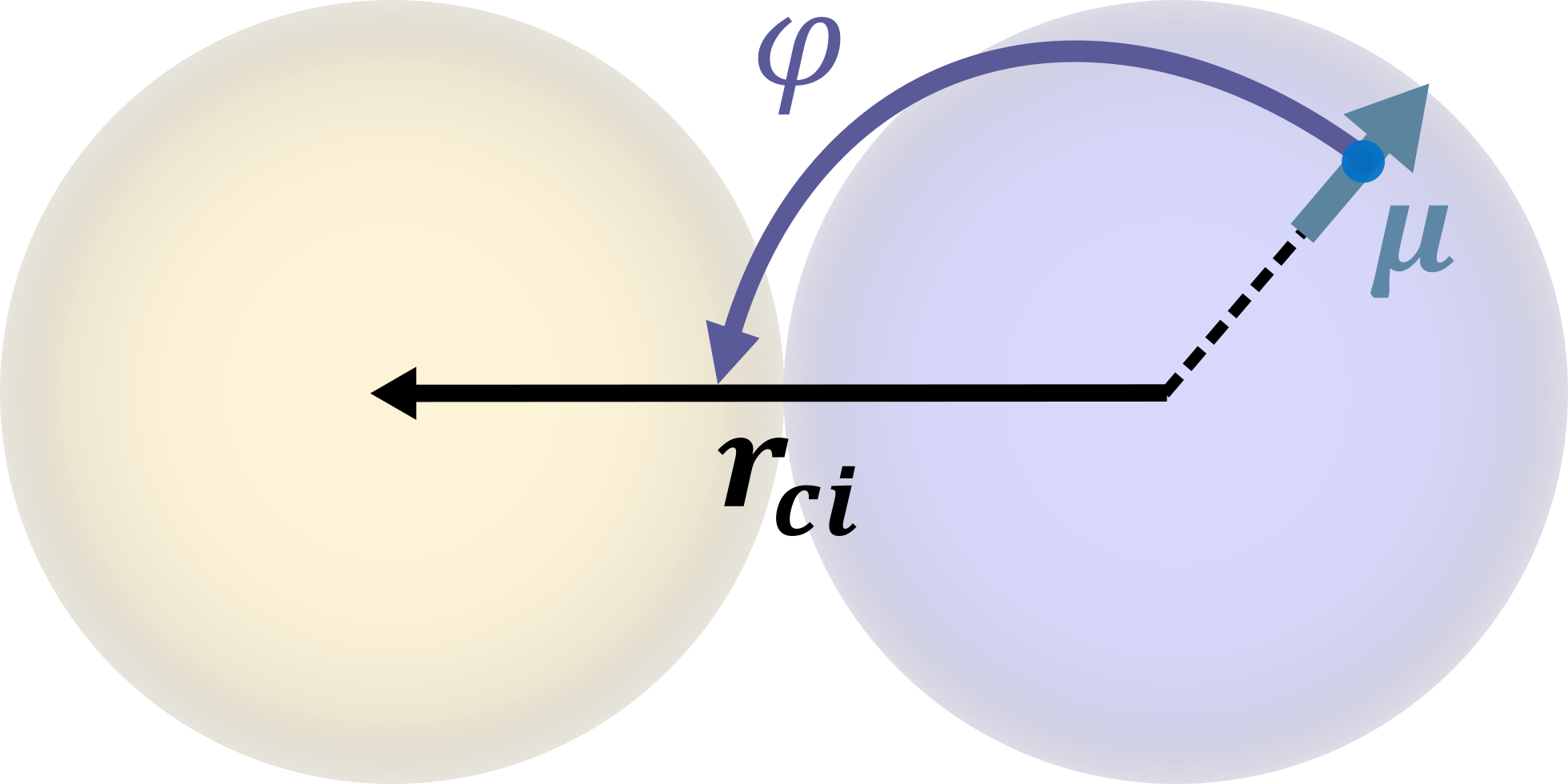}
    \caption{Schematic illustration of angular descriptors for the specific particle pair. The angles $\varphi_1$ and $\varphi_2$ denote the orientation of each dipole moment $\boldsymbol{\mu}_1$ and $\boldsymbol{\mu}_2$ relative to the dipole-displacement vector $\mathbf{r}^{d}$, while $\gamma_{12}$ is the torsional angle between the two dipoles. The scalar center-to-center distance is $r$, and $d$ represents the shift of each dipole from the particle center.}
    \label{fig:orientation}
\end{figure}

For each ion--solvent pair, we define the displacement vector $\mathbf{r}_{ci}$ to point from the solvent particle center to the ion, with scalar separation $r_{ci}=|\mathbf{r}_{ci}|$. The center--ion radial distribution function (RDF) is then:
\begin{equation}
    g_{ci}(r)
    =
    \frac{1}{4\pi r^2 \rho_s N_i}
    \left\langle
    \sum_{m=1}^{N_i}\sum_{n=1}^{N_s}
    \delta\!\left(r-r_{mn}\right)
    \right\rangle,
\end{equation}
where $N_i$ is the number of ions of the chosen species---cation or anion---$N_s$ is the number of solvent centers used in the analysis, $\rho_s$ is the solvent number density, and $r_{mn}=|\mathbf{r}_{ci}^{(n,m)}|$ is the scalar distance between ion $m$ and solvent center $n$. In this case, the simulation is dilute, with only 20 ion pairs in a 3000-particle solvent. Values of $g_{ci}(r)=1$ correspond to a spatially uniform distribution at the same average density; peaks identify preferred solvation-shell distances.

To more directly quantify the structure of the first solvation shells, it is useful to reduce each RDF to a single quantity. One appropriate scalar quantity is the coordination number $N_{\mathrm{coord}}$, which quantifies the number of solvent particles located in the first shell around a tagged ion. In these simulations, we examine one ion type in a box of solvent particles at liquid coexistence density. We obtain the coordination number by integrating the ion--solvent radial distribution function up to the location of the first minimum:
\begin{equation}
    N_{\mathrm{coord}}
    =
    4\pi \rho_s \int_0^{r_{ci,\min}} dr_{ci}\, r_{ci}^2 g_{ci}(r_{ci})~,
\end{equation}
where $r_{ci,\min}$ is the location of the first minimum following the first peak. We obtain $r_{ci,\min}$ by applying a Savitzky--Golay polynomial filter \cite{savitzkySmoothingDifferentiationData1964} to the RDF to minimize the impact of small statistical fluctuations and then locating the first local minimum after the first peak. The resulting cutoff is then applied to the original, unsmoothed RDF when evaluating $N_{\mathrm{coord}}$.

To estimate the uncertainty associated with both RDF sampling and the determination of the first-shell boundary, each trajectory is divided into blocks. An RDF is calculated independently for each block, and the first minimum $r_{ci,\min}$ is determined using the same procedure described above. Each block RDF is then integrated up to its corresponding block-specific cutoff, yielding a distribution of block coordination numbers. The reported coordination number is the mean of these block values, and its uncertainty is represented by a 95\% confidence interval obtained by resampling the block coordination numbers. The mean of the block-specific values of $r_{ci,\min}$ is used as the representative first-shell cutoff for subsequent structural analyses.

% For the unshifted-dipole system in particular, the cation and anion radial solvation structures are expected to be equivalent, up to statistical fluctuations. We therefore determine a common first-shell cutoff from the first minimum of the average cation and anion RDFs and use this shared cutoff to integrate the original, unsmoothed RDF of each ion species separately. This procedure prevents small differences in the discretized minimum location from producing artificial differences in the inferred coordination numbers.

To evaluate orientational correlations between tagged ions and surrounding solvent particles, we define the angle $\varphi$ between $\boldsymbol{\mu}$ and $\mathbf{r}_{ci}$ according to
\begin{equation}
    \cos\varphi
    =
    \frac{\boldsymbol{\mu}\cdot\mathbf{r}_{ci}}
    {|\boldsymbol{\mu}|\,|\mathbf{r}_{ci}|}.
\end{equation}
With this convention, $\cos\varphi>0$ means that the solvent dipole points toward the ion, while $\cos\varphi<0$ means that the solvent dipole points away from the ion. The corresponding angular-radial distribution function (ARDF) is the joint distribution in $r_{ci}$ and $\cos\varphi$:
\begin{equation}
    g_{ci}(r_{ci},\cos\varphi),
\end{equation}
which is the orientationally resolved analog of the ion--solvent radial distribution function. To improve statistical sampling while maintaining a dilute system, we again simulate 20 ion pairs in a 3000-particle solvent.

Equivalently, one could describe the same pair using the opposite displacement vector, $\mathbf{r}_{ic}=-\mathbf{r}_{ci}$, which points from the tagged ion to the solvent particle center. While the scalar separation is unchanged, $r_{ic}=r_{ci}$, the angular coordinate changes sign:
\begin{equation}
    \frac{\boldsymbol{\mu}\cdot\mathbf{r}_{ic}}
    {|\boldsymbol{\mu}|\,|\mathbf{r}_{ic}|}
    =
    -\frac{\boldsymbol{\mu}\cdot\mathbf{r}_{ci}}
    {|\boldsymbol{\mu}|\,|\mathbf{r}_{ci}|}
    =
    -\cos\varphi.
\end{equation}
Thus, changing the displacement-vector convention does not produce new structural information; it only reflects the ARDF about $\cos\varphi=0$. Given identical sampling and binning:
\begin{equation}
    g_{ic}(r_{ic},\cos\varphi)
    =
    g_{ci}(r_{ci},-\cos\varphi).
\end{equation}

By integrating out the radial component within the first solvation, one obtains the angular distribution function (ADF)\cite{kohnsRelativePermittivityStockmayerType2020, langenbachCoOrientedFluidFunctional2017}:
\begin{equation}
    g(\cos \varphi)=\frac{3}{4\pi r_{ci,\min}^3}\int_0^{r_{ci,\min}}dr  4 \pi r^2\,
    g_{ic}(r_{ic},\cos\varphi)\,,
\end{equation}
to isolate orientational structure. Specifically, we use the mean $r_{ci,\min}$ obtained from the block analysis described above. 

To quantify the spatial polarization of the solvation environment, we also calculate the radial solvent polarization around each ion species. Unlike the aforementioned structural distributions, as we are now interested in the charge centers, this quantity uses the dipole position $\mathbf{r}_n^d$. We define $\mathbf{r}_{di}^{(n,m)}=\mathbf{r}_n^d-\mathbf{r}_m^i$ to point outward from ion $m$ to solvent dipole $n$, with scalar separation $r_{di}^{(n,m)}=|\mathbf{r}_{di}^{(n,m)}|$. The radial polarization is evaluated in spherical shells as:
\begin{equation}
P_r(r_k)=\frac{1}{N_iV_k}
\left\langle
\sum_{m=1}^{N_i}
\sum_{\substack{n=1\\ r_{k,-}\le r_{mn}<r_{k,+}}}^{N_s}
\boldsymbol{\mu}_n\cdot\widehat{\mathbf r}_{mn}
\right\rangle,
\end{equation}
where $V_k=4\pi(r_{k,+}^3-r_{k,-}^3)/3$ is the shell volume, and the brackets denote an average over sampled trajectory frames. Here, $\widehat{\mathbf{r}}_{di}^{(n,m)}$ is the unit vector along $\mathbf{r}_{di}^{(n,m)}$, and $r_k$ is the midpoint of shell $k$. Pair displacements are evaluated using the minimum-image convention.

We use 200 uniformly spaced shells over $0<r_{di}/\sigma<5$ and sample every fourth saved trajectory frame. Positive $P_r$ denotes
net outward dipole alignment, whereas negative $P_r$ denotes net inward alignment, opposite to the sign convention for $\cos\varphi$ above. Thus, the shell-volume normalization captures both dipole alignment and local dipole density. The bound charge contribution from the dipoles can be obtained as:
\begin{equation}
    Q_\mathrm{bound}(r_k) = -4\pi r_k^2P_r(r_k)\,.
\end{equation}

A useful feature obtainable from the aforementioned properties is the ability to identify both a preferred solvation-shell distance and a preferred solvent-dipole orientation around the tagged ion. While these properties are expected to be symmetric between cations and anions for unshifted dipoles, up to reversal of dipole orientation and polarization sign, asymmetries and preferential orientations are expected to arise for shifted dipoles.

\subsubsection{Solvation Free Energies}
\label{sec:ti}

To quantify how solvent dipolar asymmetry affects ionic thermodynamics, we compute ion solvation free energies using thermodynamic integration (TI)\cite{shockSolvationEnergyIons2020,nakamuraIonSolvationLiquid2012,israelachviliIntermolecularSurfaceForces2011,eikeAtomisticSimulationSolidliquid2006}. The quantity of interest is the free-energy change associated with charging a single ion in the shifted-dipole Stockmayer solvent. Thus, our simulations use 1500 shifted-dipole Stockmayer solvent particles at coexistence density and 1 ion to avoid simultaneous cation--anion correlations. In our current model, the Lennard-Jones interactions remain fixed along the charging path; as such, TI isolates the electrostatic contribution to ionic solvation. All free energies are reported in reduced Lennard-Jones units, so $\Delta U^\ast = \Delta U/\epsilon$ and $\Delta F^\ast=\Delta F/\epsilon$. Because the simulations are performed at a reduced temperature of $T^\ast = k_\text{B} T/\epsilon = 1$, we may equivalently express $\Delta F^\ast$ as $\beta \Delta F$, which we will use for the remainder of the paper.

TI introduces a coupling parameter $\lambda$ that continuously connects two states. At $\lambda=0$, the ion carries zero charge but retains its short-range interactions. At $\lambda=1$, the ion carries its full charge and interacts electrostatically with the solvent. The coupling is implemented by scaling the ionic charge linearly with $\lambda$, with the ion--solvent and ion--ion short-range interactions are held fixed, following the charging protocol described in previous studies~\cite{reifCorrectionEffectiveElectrostatic2015, linIonicSolvationStudied2011}.

The potential energy thus depends on $\lambda$, denoted by $U^\ast(\lambda)$, and the free-energy difference between the two endpoints is obtained from:
\begin{equation}
    \beta \Delta F = \int_0^1 
    \left\langle \frac{\partial U^\ast(\lambda)}{\partial \lambda} \right\rangle_{\lambda}
    \, d\lambda,
\end{equation}
where $\langle \cdots \rangle_{\lambda}$ denotes an equilibrium ensemble average taken in a simulation at fixed $\lambda$. 

% In practice, the integral is evaluated from a discrete set of intermediate $\lambda$ windows. At each value of $\lambda$, a molecular dynamics simulation is performed and the ensemble average
% \begin{equation}
%     \left\langle \frac{\partial U^\ast(\lambda)}{\partial \lambda} \right\rangle_{\lambda}
% \end{equation}
% is accumulated. The derivative $\partial U^\ast/\partial \lambda$ is the TI integrand; since $\lambda$ is dimensionless, it is reported in units of $\epsilon$. It describes how rapidly the potential energy changes as the ionic charge is increased along the charging path. Although $\lambda$ is fixed during a given simulation, $\partial U^\ast/\partial \lambda$ is evaluated as a configuration-dependent observable: it measures how the energy of the same instantaneous configuration would change under an infinitesimal change in $\lambda$. The free-energy estimator built into LAMMPS evaluates this quantity along the trajectory at fixed $\lambda$. Averaging over each charging window yields the TI integrand, and numerical integration over $\lambda$ then yields the charging free energy. TI therefore replaces a direct comparison between the endpoint states with the integration of a quantity that can be measured continuously along the charging path.

For each ion species and solvent parameter set, the TI integral is evaluated using 18 fixed-$\lambda$ windows spanning the charging path from $\lambda=0$ to $\lambda=1$. Each window is equilibrated for $3\times 10^4$ timesteps and then run for $7\times 10^5$ production timesteps at $T^\ast=1.0$, with trajectory and TI quantities sampled every $100$ timesteps. This procedure was tested by monitoring the convergence of the potential energy and the accumulated averages of $\left\langle \partial U^\ast/\partial\lambda \right\rangle_\lambda$ with simulation time. The final charging free energy is obtained by numerically integrating the resulting TI curve over $\lambda$. Uncertainties are estimated by block averaging the TI integrand within each $\lambda$ window and propagating the resulting standard errors through the numerical integration.

Conceptually, a more negative value of $\beta \Delta F$ means that a given ion is solvated more strongly. As a result, comparing cation and anion charging free energies can show how strongly solvent dipolar asymmetry distinguishes positive and negative ions. 

\subsubsection{Potential of Mean Force}
To examine how the solvent dipole magnitude and displacement alter the effective interaction between oppositely charged ions, we calculate the ion-pair potential of mean force (PMF) using the adaptive biasing force (ABF) method\cite{darveAdaptiveBiasingForce2008,heninExploringMultidimensionalFree2010} implemented in the Colvars module \cite{fiorinUsingCollectiveVariables2013}. The calculations contain one cation--anion pair and 5000 shifted-dipole Stockmayer solvent particles, thereby approximating the infinite-dilution limit while maintaining overall charge neutrality. The size of the simulation box is adjusted for each solvent parameter set to reproduce the corresponding coexisting liquid density obtained in our previous work\cite{walkerStockmayerFluidShifted2026a,varnerStockmayerFluidShifted2026}. Consequently, changes in the PMF incorporate both the direct effect of the solvent dipolar parameters and their associated effect on the equilibrium liquid density.

The collective variable used in the ABF calculations is the separation between the centers of the cation and anion,
\begin{equation}
    r = \left|\mathbf{r}_{+}-\mathbf{r}_{-}\right|,
\end{equation}
which is sampled continuously over the interval $0.75\sigma \leq r \leq 8\sigma$.
The lower bound permits sampling of the contact ion-pair configuration while avoiding separations that are effectively inaccessible because of the steep Lennard-Jones repulsion. The upper bound is chosen to reach the large-separation region in which the solvent-mediated ion--ion interaction approaches its asymptotic value.

The PMF, $W(r)$, is the reversible free-energy change associated with bringing the two ions from a reference separation to a distance $r$. It is related to the equilibrium radial distribution function in the dilute-pair limit by
\begin{equation}
    \beta W(r) = -\ln g_{+-}(r) + C,
\end{equation}
where $C$ is an arbitrary additive constant. Equivalently, the derivative of the PMF is the negative of the mean force acting along the ion-separation coordinate:
\begin{equation}
    \frac{\mathrm{d}W(r)}{\mathrm{d}r}
    =
    -\left\langle F_r \right\rangle_r,
\end{equation}
where $\langle F_r\rangle_r$ denotes a conditional ensemble average at fixed ion separation. The radial-coordinate Jacobian is accounted for by the ABF implementation, so the resulting $W(r)$ represents the interaction PMF associated with $g_{+-}(r)$, rather than the uncorrected free energy obtained directly from the probability distribution of the scalar distance.

% Direct sampling of $W(r)$ is inefficient when the ion pair must cross large free-energy barriers or escape strongly bound configurations. ABF addresses this problem by dividing the reaction-coordinate interval into spatial bins and accumulating an estimate of the mean force within each bin. As this estimate converges, an opposing biasing force is applied to cancel the local mean force and promote approximately uniform sampling across the full separation range. The bias therefore facilitates transitions between contact, solvent-separated, and dissociated ion-pair configurations without prescribing the shape or locations of these states. The unbiased mean-force profile accumulated during the simulation is then numerically integrated along $r$ to reconstruct the PMF:
% \begin{equation}
%     \beta W(r)
%     =
%     -\beta\int_{r_{\mathrm{ref}}}^{r}
%     \left\langle F_{r'}\right\rangle_{r'}
%     \,\mathrm{d}r'.
% \end{equation}
% Because only free-energy differences are physically meaningful, the additive constant is chosen such that
% \begin{equation}
%     \beta W(8\sigma)=0.
% \end{equation}
% Negative values of $\beta W(r)$ therefore indicate that bringing the ions to a separation $r$ is thermodynamically favorable relative to the largest sampled separation, whereas positive values indicate an unfavorable configuration.

The system is first equilibrated for $2\times10^{6}$ timesteps, followed by $1\times10^{7}$ production timesteps, during which the mean force and sampling count are accumulated as functions of the ion separation. Convergence is assessed by monitoring the sampling frequency of the reaction-coordinate bins and the relative change in the accumulated mean-force profile with simulation time. The converged mean-force profile is integrated numerically and shifted to satisfy $W(8\sigma)=0$.

Minima in the resulting PMF identify favorable ion-pair configurations. A minimum near $r\approx\sigma$ corresponds to a contact ion pair, whereas minima at larger separations correspond to configurations in which one or more solvent layers separate the ions. The depths of these minima relative to the large-separation reference quantify the stability of the corresponding ion-pair states. Comparing the cation--anion PMFs across dipole magnitudes and displacements therefore reveals how solvent polarity and dipolar asymmetry modify ion association and the relative stability of contact and solvent-separated configurations.

\section{Results and Discussions}
We now examine the solvation structure of the shifted-dipole Stockmayer fluid around cations and anions. On the basis of these results, we subsequently analyze the solvation free energies of single ions and potentials of mean force between oppositely charged ions.

Throughout this section, reduced Lennard-Jones units are employed. Accordingly, we define the dimensionless density, dipole moment, and shift as:
   \begin{equation}
    \rho^*=\rho\sigma^3,\quad
    \mu^* = \frac{\mu}{\sqrt{4\pi\epsilon_0\epsilon\sigma^3}},\quad
    d^*=\frac{d}{\sigma},\nonumber
\end{equation} 
\begin{equation}
     T^*=\frac{kT}{\epsilon},\quad q^* =
\frac{q}{\sqrt{4\pi\epsilon_0\epsilon\sigma}},\nonumber
\end{equation}

where the superscript $*$ denotes reduced quantities. All simulations use $T^*=1$. As a reference, if $\epsilon/k=300\,\mathrm{K}$ and $\sigma=0.275\,\mathrm{nm}$, then $\mu^*\approx1.95$ and $\rho^*=0.9$ correspond approximately to liquid water near room temperature at $T^*=1$.

\subsection{Structural Properties}
\label{sec:struct}

\begin{figure}
    \centering
    \includegraphics[width=\linewidth]{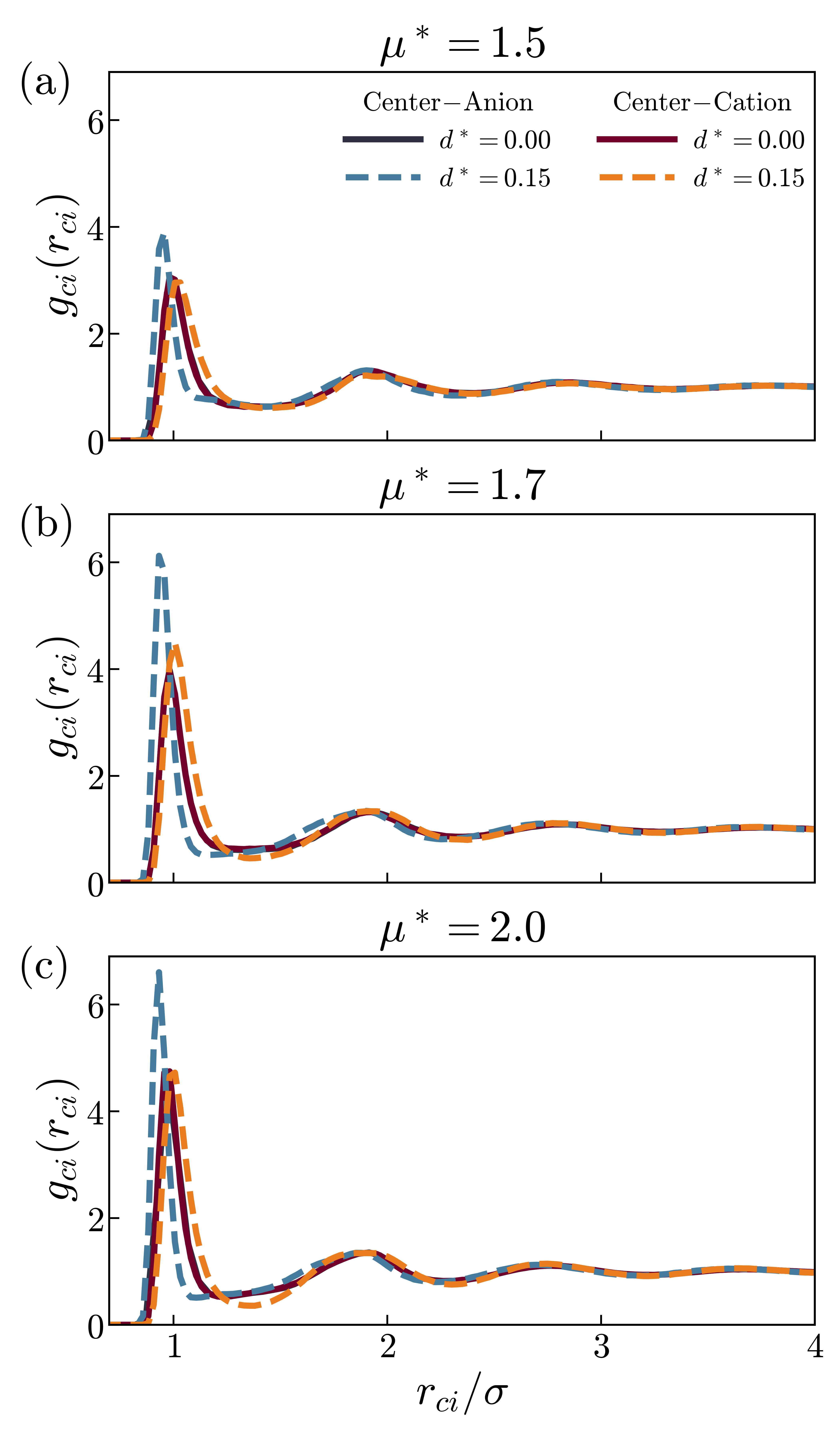}
    \caption{Solvent center--ion radial distribution functions for anions and cations at $\mu^\ast=1.0$, $1.5$, and $2.0$. Solid curves correspond to $d^\ast=0.00$, while dashed curves correspond to $d^\ast=0.15$.}
    \label{fig:rdfs}
\end{figure}

\begin{figure}
    \centering
    \includegraphics[width=\linewidth]{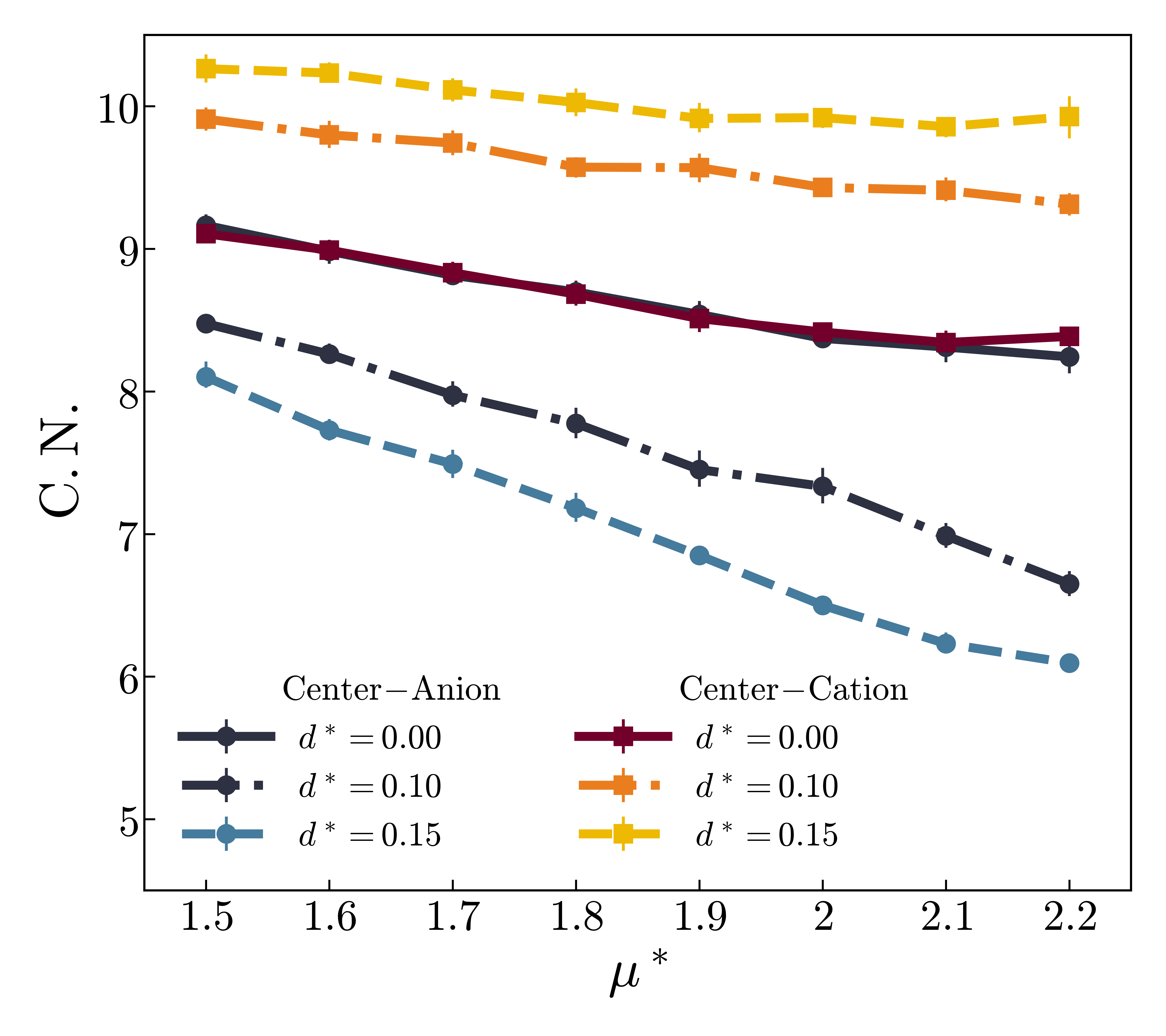}
    \caption{Coordination numbers of solvent particles around anions and cations as functions of dipole strength, obtained by integrating the corresponding radial distribution functions over the first solvation shell up to the mean first-shell cutoff \(r_{\mathrm{cut}}\). Error bars represent 95\% confidence intervals obtained by resampling block-averaged coordination numbers.}
    \label{fig:coord_num}
\end{figure}

% for fig 4, find error in r_min: a min and max value for each. then do the integration of C.N.
% also try constant r_min = 1.5 cut-off

In panels (a)--(c) of Fig.~\ref{fig:rdfs}, we show the solvent--ion RDFs, which describe how solvent molecular centers are distributed around a tagged ion. When the dipole is unshifted (i.e., $d^* = 0.00$), the cation and anion curves, shown as solid lines, are identical for all three dipole strengths. This is to be expected as the ion--dipole interaction strength is expected to be perfectly symmetric. Between dipole strengths, we recover the expected trend that the first peak, corresponding to the location of the first solvation shell, intensifies, as the ion--dipole interaction strengthens\cite{shockMolecularDynamicsSimulations2023,shockSolvationEnergyIons2020}. The subsequent, smaller peaks correspond to the weaker solvation shells at greater distances.

Upon introducing a dipole shift of $d^\ast=0.15$, shown by the dashed curves, the first peaks in the RDFs visibly separate for the cation and anion. In the case of the anion, the first peak shifts to lower $r_{ci}$ and intensifies strongly. This can be rationalized by the fact that the head of the dipole is now more accessible to the dipole, resulting in a stronger ion--dipole interaction. By a similar token, the first peak in the RDF for the cation shifts to larger $r_{ci}$ at the tail of the dipole is now less accessible, resulting in a weaker ion--dipole interaction. Note that outwards shift in the peaks for the cations isn't accompanied by a weakened peak. This is likely because, with increasing dipole shift, the density of the system also increases\cite{walkerStockmayerFluidShifted2026a}, compensating for the weaker interaction. Another interesting observation is that, beyond the first peak, all higher order peaks remain at largely the same location and intensity, for a given dipole strength. This seems to indicate that the effects of dipolar asymmetry are most pronounced within the first solvation shell only. This was also found to be the case in the pure solvent systems\cite{walkerStockmayerFluidShifted2026a}.

\begin{figure*}
    \centering
    \includegraphics[width=\linewidth]{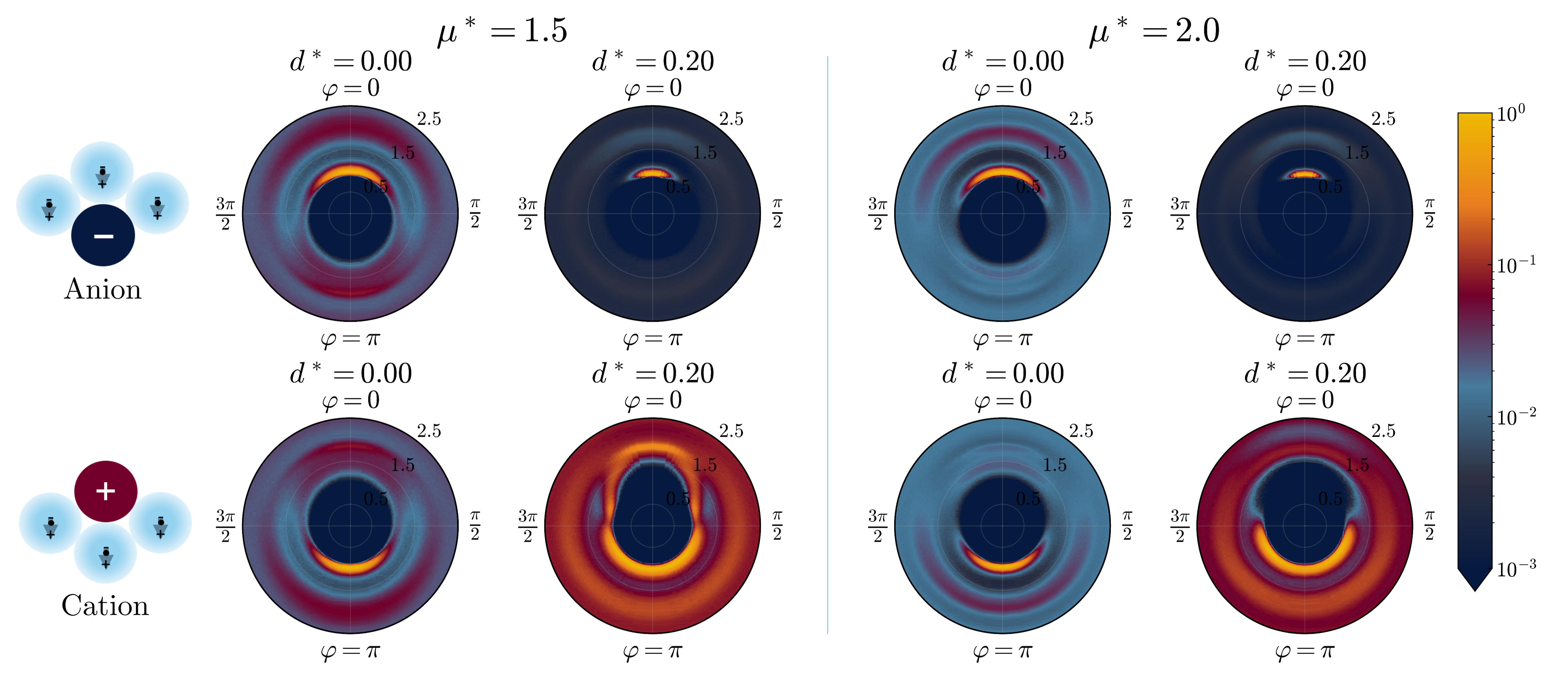}
    \caption{Circular representation of the ion--solvent angular-radial distribution functions. Radius denotes the ion--solvent-center separation $r_{ci}/\sigma$ up to 2.5, and $\varphi$ denotes the angle between the solvent dipole and the displacement vector from the solvent center to the ion (Fig.~\ref{fig:orientation}). Color denotes the logarithmically scaled, normalized probability density, so brighter regions correspond to more probable ion–solvent configurations. The schematics on the left illustrate the information contained in the ARDFs, depicting the preferred solvent orientation and the regions around a given ion where solvent particles are most likely to be found.}
    \label{fig:ardfs}
\end{figure*}

% radial polarization profile: dipole distribution and differentiate, similar operation as E over interface. integrating should cancel the scalar charge. pick one case unshifted, other shifted; anion cation as well
% we have ardfs of solvent around ion, we need ardfs of ion around solvent: bin the dipole orientations cos(phi) --> avg cos(phi) over r

As shown in Fig.~\ref{fig:coord_num}, coordination numbers at different combinations of $\mu^\ast$ and $d^\ast$ capture the same distinction shown in Fig.~\ref{fig:rdfs}. At $d^\ast=0.00$, cation and anion coordination numbers remain identical over the full range of dipole strengths. With increasing dipole strength, the coordination number decreases. This is to be expected as, although the ion--dipole interaction of a single pair increases, interactions between dipoles within the first solvation shell become more unfavorable. As such, it is preferable to have fewer solvent particles within the first solvation shell. 

At finite shifts, the curves corresponding to cations and anions once again separate. The coordination number for the cation noticeably increases with dipole shift. This may seem unexpected as the ion--dipole interaction is now weaker. However, examining Fig.~\ref{fig:rdfs}, we note that the peak intensity between the unshifted and shifted cases remain the same, only moving to higher $r_{ci}$ values. As the coordination number weighs the RDF by $r_{ci}^2$, this shift implies that, as the solvent particles are now preferentially located further away, more of them can fit within the first solvation shell. In addition, as the shifted dipoles are now pointing away from the central ion, the dipole--dipole repulsions are also weakened, resulting in a slower decrease of coordination number with increasing $\mu^*$.

The opposite is true for the case of the anion coordination number. While the first peak has intensified, its location at lower $r_{ci}$ means that fewer solvent particles can fit within the first solvation shell. The stronger dipole--dipole repulsion within the first solvation shell also account for the faster decrease in coordination number with increasing $\mu^*$.

\begin{figure}
    \centering
    \includegraphics[width=1\linewidth]{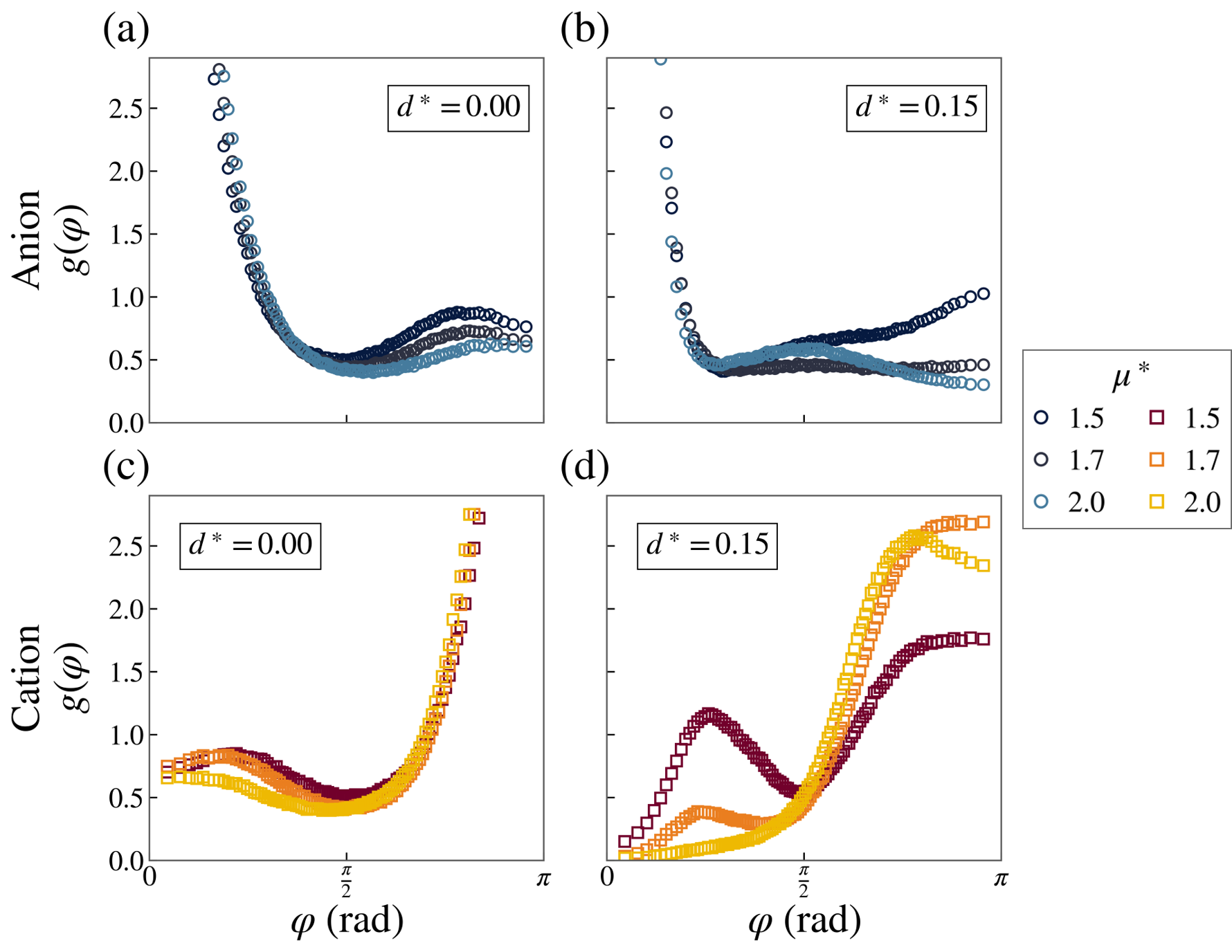}
    \caption{Ion--solvent angular distribution functions obtained by integrating the ARDF over the first solvation shell. The integration limit is the corresponding mean first-shell cutoff $r_{\mathrm{cut}}$ for each state point. The distributions are normalized such that
    $g(\varphi)=1$ corresponds to isotropic solvent orientation.}
    \label{fig:adfs}
\end{figure}

While the metrics reported thus far help analyze the radial distribution of solvent particles around the central ions, they do not resolve differences in the orientational structure. We therefore examine a polar representation of the ion--solvent angular--radial distribution functions (ARDFs) in Fig.~\ref{fig:ardfs}, which describes the joint dependence of the solvent distribution on the center--ion separation $r_{ci}$ and the angle $\varphi$ between the solvent dipole and the solvent-to-ion displacement vector. We also provide representative angular distribution functions (ADFs) in Fig.~\ref{fig:adfs} to isolate the angular-dependence.

For the unshifted solvent systems, we note that the ARDFs and ADFs are completely symmetric around $\varphi=\frac{\pi}{2}$ between the anion and the cation, with the preferential angle being $\varphi=0$ for the anion and $\varphi=\pi$ for the cation, corresponding to the orientations with the strongest ion--dipole interaction. With increasing dipole strength, these regions only intensify. We do note, within the ADFs, the presence of peaks around $\varphi=\frac{\pi}{4}$ for the anions and $\varphi=\frac{3\pi}{4}$ for the cations in Figs.~\ref{fig:adfs}(a) and \ref{fig:adfs}(c) for low dipole strengths. These peaks likely correspond to solvent particles present within the first solvation shell, which primarily interact favorably with other dipoles whose orientations face the central ion. As the dipole strength increases and the size of the first solvation shell decreases, these dipoles are ejected.

In the shifted dipole cases, around anions, the dominant first-shell feature remains centered near $\varphi\approx0$ and $r_{ci}/\sigma\approx1.0$ for all conditions shown. The primary change is that the size of this region is reduced and it becomes significantly more favorable to have a $\varphi=0$ orientation around the ion. This reflects the strengthening ion--dipole interaction but also highlights how restricted the dipole orientation around the anion becomes due to the shift. This is also apparent in Fig.~\ref{fig:adfs}(b) where there is now a sharp increase in the ADF at low $\varphi$.

Around cations, the distribution remains symmetric around $\varphi=\pi$ but widens for finite shifts such as $d^*=0.2$. This likely occurs because the shifted dipole interacts most strongly with the ion at an angle that is offset from $\varphi=\pi$. At $\mu^\ast=2.0$, this region becomes somewhat more localized, although it remains broader than the head-on feature observed around the anion. These wider distributions reflect the less-restricted orientations of the solvent particles around the central cation, as seen in Fig.~\ref{fig:adfs}(d), where the ADF flattens at larger $\varphi$.

\begin{figure}
    \centering
    \includegraphics[width=\linewidth]{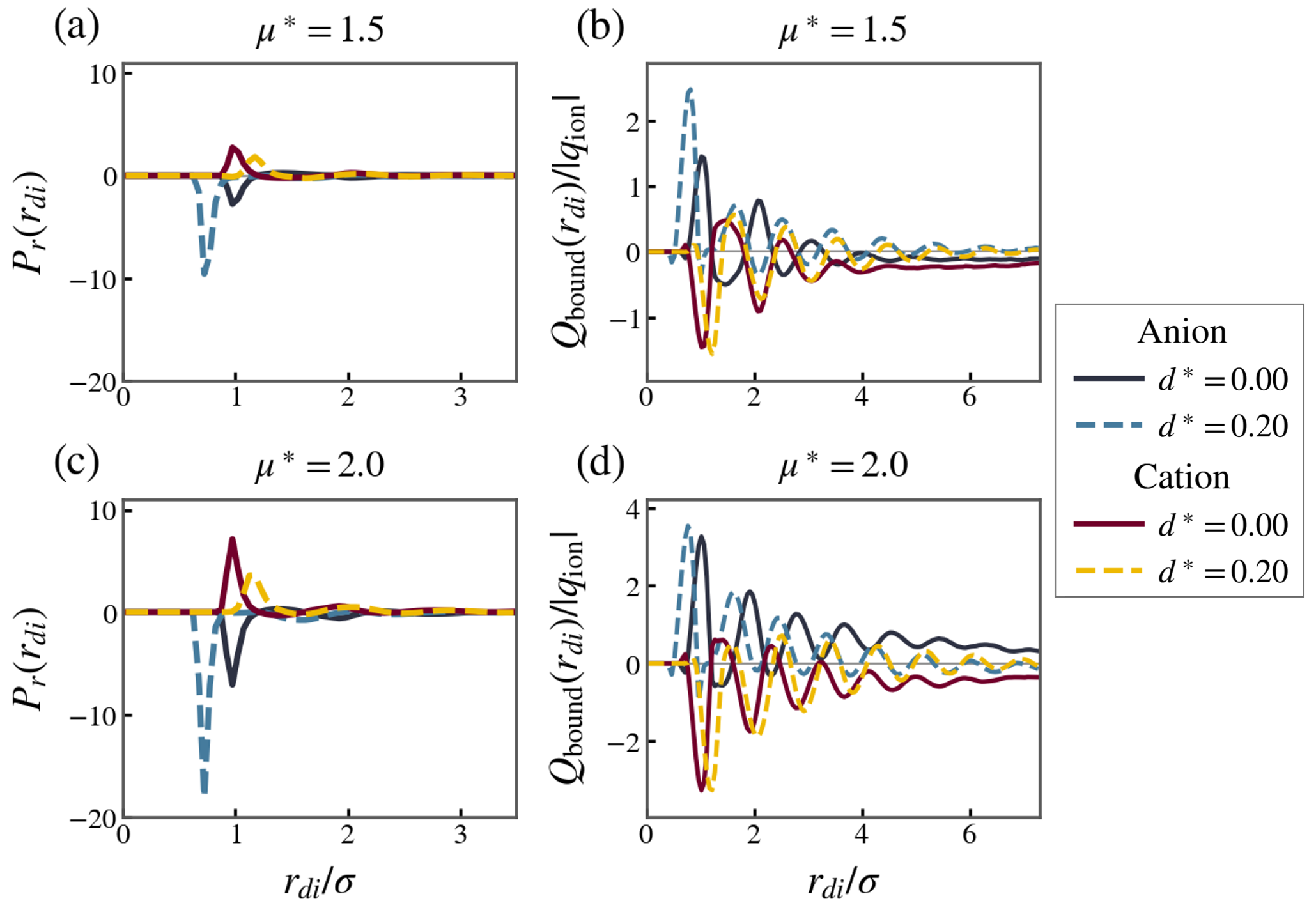}
    \caption{Radial solvent polarization around anions and cations for (a) $\mu^\ast=1.5$ and (c) $\mu^\ast=2.0$, comparing $d^\ast=0.00$ and $0.20$. We also include the bound charge contributed from the dipoles in (b) for $\mu^\ast=1.5$ and (d) for $\mu^\ast=2.0$.}
    \label{fig:radpol}
\end{figure}

To assess how the preferential orientation of solvent dipoles contributes to the spatial polarization around ions, we calculate the radial solvent polarization, $P_r(r_{di})$, and the resulting bound charge, $Q_\mathrm{bound}(r_{di})$, shown in Fig.~\ref{fig:radpol}. For unshifted dipoles, the polarization around anions and cations is equal in magnitude and opposite in sign, regardless of dipole strength, consistent with our previous analysis (Fig.~\ref{fig:radpol}(a) and (c)). Shifting the dipoles breaks this symmetry. Because shifted dipoles can approach anions more closely, they produce stronger polarization at small $r_{di}$. Around cations, the dipole centers remain farther from the ions, and the broader orientational distribution reduces the radial projection, producing a weaker outward polarization. Increasing $\mu^*$ strengthens the response through both the larger dipole moment and the stronger ion--dipole coupling, while preserving this asymmetry.

To assess how the preferential orientation of solvent dipoles contributes to the spatial polarization around ions, we calculate the radial solvent polarization, $P_r(r_{di})$, and the resulting bound charge, $Q_\mathrm{bound}(r_{di})$, shown in Fig.~\ref{fig:radpol}. For unshifted dipoles, the polarization around anions and cations is equal in magnitude and opposite in sign, regardless of dipole strength, consistent with our previous analysis (Fig.~\ref{fig:radpol}(a) and (c)). Shifting the dipoles breaks this symmetry. Because shifted dipoles can approach anions more closely, they produce stronger polarization at small $r_{di}$. Around cations, the dipole centers remain farther from the ions, and the broader orientational distribution reduces the radial projection, producing a weaker outward polarization. Increasing $\mu^*$ strengthens the response through both the larger dipole moment and the stronger ion--dipole coupling, while preserving this asymmetry.

The resulting bound-charge contribution is shown in Fig.~\ref{fig:radpol}(b) and (d). The stronger polarization of shifted dipoles around anions produces substantially greater screening within the first solvation shell than in the unshifted case. This occurs despite shifted-dipole systems generally having lower dielectric constants than their unshifted counterparts \cite{walkerStockmayerFluidShifted2026a,langenbachRelativePermittivityDipolar2020} (see SI for numerical values). Around cations, the shifted dipoles produce weaker polarization than the unshifted dipoles; however, because the first solvation shell is larger, the resulting screening is comparable in the two cases. The oscillations in the polarization and bound charge reflect the ordered packing of dipoles caused by excluded-volume interactions at high densities. At sufficiently large separations, the bound charge is expected to approach zero.\footnote{These systems contain multiple ions, so the contribution of the free ions to the screened charge is expected to cancel the charge of the central ion.} Note that these oscillations decay much more slowly within the unshifted dipole cases than they do in the shifted, particularly in Fig.~\ref{fig:radpol}(d). This reflects how, at longer ranges, the dielectric screening of the solvent dominates, resulting in the higher dielectric solvent (the unshifted dipoles) having a larger $Q_\mathrm{bound}(r)$.\footnote{For interested readers, we provide an analytical theory to corroborate these results within the supporting information.}
\subsection{Solvation Free Energies}

With a more complete understanding of the solvent microstructure around isolated ions, we now investigate the solvation thermodynamics. We do so by computing ionic solvation free energies with the thermodynamic integration procedure described in Sec.~\ref{sec:ti}. Along the described charging path, $\beta \Delta F$ isolates the electrostatic part of the solvation free energy and so directly measures how strongly the solvent distinguishes positive and negative ions. A sign-dependent solvation response of this kind is well established in existing literature on charge hydration asymmetry, where ions of opposite sign but comparable size can have different solvation free energies and distinct local solvent structures \cite{rajamaniSizeDependentIon2004,grossfieldDependenceIonHydration2004,mobleyChargeAsymmetriesHydration2008,mukhopadhyayChargeHydrationAsymmetry2012,oneillPairNotPair2024}.

The ionic solvation free energies are plotted against the dimensionless dipolar coupling parameter $\xi=\frac{\rho \mu^2}{3\epsilon_0 k_{\mathrm B}T}$, which combines the solvent number density, dipole magnitude, and thermal energy into a single measure of the overall dipolar interaction strength. In terms of reduced quantities in Lennard-Jones units, the coupling parameter becomes:
\begin{equation}
    \xi = \frac{4\pi\rho^\ast(\mu^\ast)^2}{3T^\ast}\,,
\end{equation}
using the coexistence liquid density corresponding to each value of $\mu^\ast$ and $d^\ast$. Plotting against $\xi$, rather than $\mu^\ast$ alone, therefore accounts for the changes in liquid density that accompany changes in both dipole magnitude and displacement.

\begin{figure}[ht!]
    \centering
    \includegraphics[width=\linewidth]{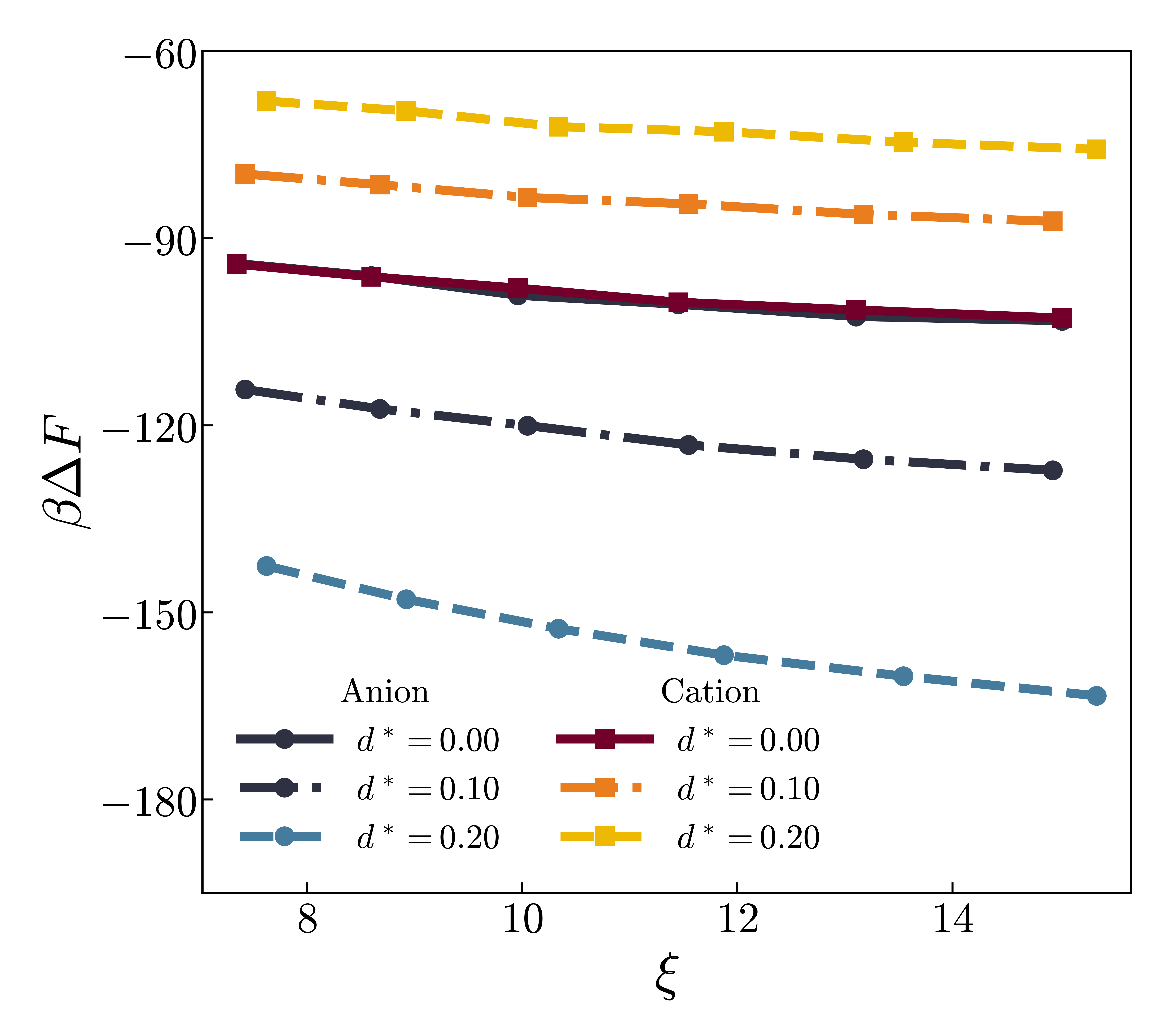}
    \caption{Ionic solvation free energies $\beta\Delta F$ as functions of the dimensionless dipolar coupling parameter $\xi$ for cations and anions in the shifted-dipole Stockmayer fluid. Color distinguishes ion sign and dipole shift $d^\ast$.}
    \label{fig:solvation_free_energy}
\end{figure}

% The resulting ionic solvation free energies are shown in Fig.~\ref{fig:solvation_free_energy}, with magnitudes that are consistent with prior works\cite{varnerEntropicOriginIonic2023,shockSolvationEnergyIons2020} and are large due to the fact we have treated ions as bare charges, neglecting the effects of polarizability.

In Fig.~\ref{fig:solvation_free_energy}, we show that the solvent with unshifted dipoles gives identical solvation free energies for cations and anions across the range of $\xi$ shown, consistent with the perfectly symmetric solvation structure. Once the dipole is displaced, however, changes in the solvation free energy between the oppositely charged ions respond in opposite directions. In increasing $d^\ast$, the cation solvation free energy decreases in magnitude, while the anion free energy deepens substantially; the same geometric asymmetry that broke the symmetry in the distribution functions and polarizability discussed in the previous section now breaks the thermodynamic symmetry between positive and negative ions. Note that these findings are also consistent with the sign effect seen in earlier simulation studies\cite{grossfieldDependenceIonHydration2004,mobleyChargeAsymmetriesHydration2008} showing that polar solvents can favor one ion sign over the other even when the solutes differ only in charge sign.

These results may initially seem inconsistent with the trends in coordination numbers observed in Fig.~\ref{fig:coord_num}, where cations have a larger coordination number compared to the anions. However, the stronger solvation free energy for anions arises from the fact that, while there are slightly fewer solvent particles around the central ion, the dipole shift results in a stronger ion--dipole interaction per solvent molecule. 

Much like the coordination number, the dependence on dipolar coupling, $\xi$, also differs for the two ion species, for very similar reasons. In the case of the cations, increasing $\xi$ lowers $\beta \Delta F$ only modestly, and the variation becomes weaker as the dipole shift increases. This is due to the inherently weaker ion-dipole interactions, resulting in a larger first solvation shell which doesn't change significantly with increasing $\xi$. As for the anions, with larger $d^\ast$, increasing $\xi$ drives a much faster decrease in $\beta \Delta F$. The more compact nature of the first solvation shell around anions and the stronger ion-dipole interaction arising due to the dipole asymmetry inherently make the solvation free energy more sensitive to the dipolar coupling, $\xi$.

Another interesting approach to examine these results is through the classical Born picture\cite{bornUeberAbsoluteBerechnung1988}, where the solvation free energy of a given ion should obey the following relationship:
\begin{equation}
    \Delta G_{\mathrm{Born}}=-\frac{q^2}{8\pi\epsilon_0\sigma_{\mathrm{Born}}}\left(1-\frac{1}{\epsilon_\mathrm{r}}\right)\,,
\end{equation}
where $\epsilon_\mathrm{r}$ is the dielectric constant of the medium and $\sigma_{\mathrm{Born}}$ is the Born solvation diameter of the ion. As shown in our previous work\cite{walkerStockmayerFluidShifted2026a}, with increasing dipole shift, $\epsilon_r$ is expected to decrease. As such, assuming all other variables remain constant, with increasing dipole shift, $\Delta G_{\mathrm{Born}}$ is expected to weaken. However, as demonstrated in Fig.~\ref{fig:solvation_free_energy}, this is only the case for the cation. In addition, attempting to project $\Delta F$ on $\left(1-1/\epsilon_\mathrm{r}\right)$ did not result in any universal collapse for either ion. This reflects the significant changes in the structure of the first solvation shell that arise from the dipolar shift and lead to different behavior across cations and anions. Thus, the effect of dipole asymmetry changing the solvation structure around ions can therefore be interpreted as increasing the Born solvation diameter for cations, and decreasing it for anions. This interpretation is consistent with the results observed in the previous section.\footnote{For interested readers, we provide an analytical theory to corroborate these results within the supporting information.}

\subsection{Potential of Mean Force}

\begin{figure}[ht!]
    \centering
    \includegraphics[width=0.9\linewidth]{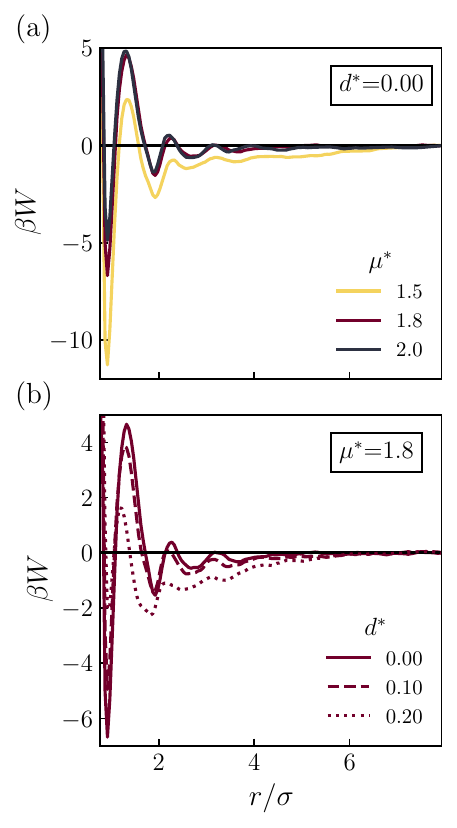}
    \caption{Potential of mean force, $\beta W$, as a function of cation--anion separation with varying dipole strength (a) and dipole shift (b). Color distinguishes dipole magnitude, and linestyle distinguishes dipole shift $d^\ast$.}
    \label{fig:pmf_r}
\end{figure}

While the prior sections have focused on the solvation structure around isolated ions, the changes in solvation free energy being in opposite directions for each ion motivate further investigation into how the asymmetric structure alters the effective attraction between oppositely charged ions. Under the classical effective Coulomb interaction in an implicit solvent, where the potential is inversely proportional to $\epsilon_\mathrm{r}$, ionic interactions should only become stronger with increasing dipole shift $d^\ast$. This is a result of depression of the dielectric constant with a shifted dipole, as shown in our previous work~\cite{walkerStockmayerFluidShifted2026} and within the SI. To quantify the actual effect of dipole asymmetry of the solvent on the effective attraction between oppositely charged ions, we calculate the cation--anion potential of mean force (PMF), shown in Fig.~\ref{fig:pmf_r}. The PMF is referenced to zero at a large separation, such that negative values correspond to favorable ion-pair configurations.

In all cases examined, the PMF exhibits a short-range minimum near contact, followed by a positive maximum and oscillations of decreasing magnitude at larger separations. The first minimum thus characterizes the stability of the contact ion pair after overcoming the first solvation shell. The subsequent oscillation represents the free-energy barrier associated with subsequent solvation shells.

At fixed $d^\ast = 0.00$ in Fig.~\ref{fig:pmf_r}(a), increasing $\mu^\ast$ reduces the magnitude of the first minimum. This can be understood due to the stronger solvent dipoles being more effective at shielding long-range ion--ion interactions, as reflected in their larger dielectric constant\cite{varnerEntropicOriginIonic2023,walkerStockmayerFluidShifted2026a,langenbachRelativePermittivityDipolar2020}.

However, examining the effect of introducing dipole asymmetry at fixed $\mu^\ast = 1.8$, shown in Fig.~\ref{fig:pmf_r}(b), we find that the overall effect is somewhat more complicated. Examining the magnitude of the first minimum, we find that it becomes less negative with increasing $d^\ast$. From the classic Coulombic picture, this is unexpected as the lower dielectric constant of shifted dipolar solvents should strengthen the attraction between oppositely charged ions. However, considering the significant changes in the first solvation shell when introducing dipole asymmetry discussed in the prior sections, we know that the solvent particles are more tightly packed around the anion. As such, to bring the cation in contact with the anion, we must disrupt the tightly bound solvent particles, incurring a larger penalty for forming the pair. The reason this barrier isn't compensated by the weaker binding around the cation is primarily because the two effects are not commensurate; the weaker solvation around the cation does not compensate for the stronger solvation around the anion. With increasing dipole shift, the magnitude of this first minimum reduces further, to the extent that the first minimum becomes shallower than the second. Such behavior is typical in the PMFs of real salts in water\cite{oneillPairNotPair2024} and highlights how the asymmetries introduced in the sd-SF model reflect real system behavior.

Interestingly, at separations beyond the first solvation shell, the trend in the interaction strength from the PMFs for varying dipole shift reverses. This is consistent with the expected behavior from the effective Coulomb interaction potential where the decrease in dielectric constant with increasing dipole shift strengthens attractions between oppositely charge ions. This aligns with the expectation that long-range charge interactions are governed not by short-range ion--dipole correlations, but rather the macroscopic screening behavior of the solvent encoded.

\begin{figure*}[ht!]
    \centering
    \includegraphics[width=\linewidth]{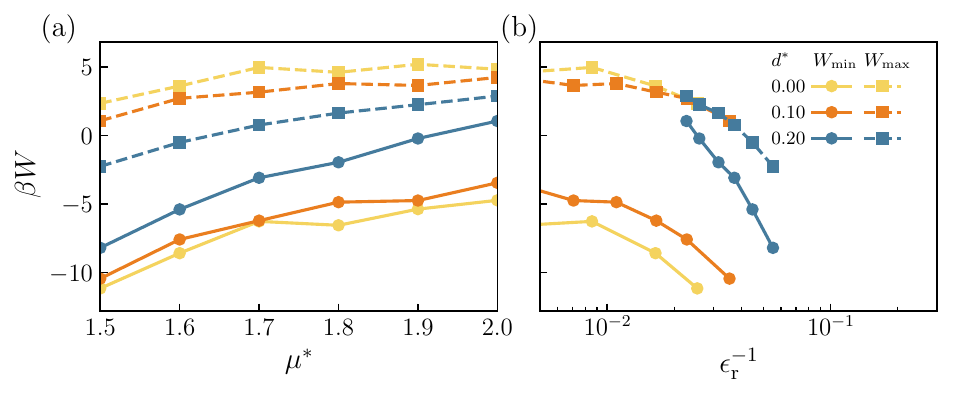}
    \caption{First minimum ($\beta W_\mathrm{min}$, solid curve, circular markers) and first maximum ($\beta W_\mathrm{max}$, dashed curve, square markers) of the potential of mean force as a function of dipole strength (a) and dielectric constant (b) with varying dipole shift. Color distinguishes dipole shift.}
    \label{fig:pmf_max_min}
\end{figure*}

These trends are summarized in Fig.~\ref{fig:pmf_max_min} by considering only the value of the PMF at its first minimum ($\beta W_\mathrm{min}$) and its first maximum ($\beta W_\mathrm{max}$). As shown in Fig.~\ref{fig:pmf_max_min}(a), for every dipole shift, $\beta W_\text{min}$ and $\beta W_\text{max}$ both increase with $\mu^\ast$, highlighting the improving ability of the solvent to shield cation--anion attractions. However, the effect of the dipole shift results in an increasing $\beta W_\mathrm{min}$ and a weakening $\beta W_\mathrm{max}$, reflecting the differences in the effects of dipole asymmetry at short- and long-ranges.

In Fig.~\ref{fig:pmf_max_min}(b), we project the results from Fig.~\ref{fig:pmf_max_min}(a) using the inverse dielectric constant (given in SI), thereby expressing them in terms of the electrostatic coupling variable that appears in the effective Coulomb potential within an implicit solvent. Note that, for $\beta W_\mathrm{min}$, the curves for different $d^\ast$ do not collapse onto a universal curve, demonstrating that the local ion--ion interaction is not solely determined by the bulk dielectric response. Rather, solvents with comparable dielectric shielding can produce different ion-pair stabilities and separation barriers depending on the asymmetries of the molecular dipole. On the other hand, at longer ranges ($\beta W_\mathrm{max}$), a universal collapse is observed between the different dipole shifts, highlighting that beyond the first solvation shell, the effects of dipole asymmetry can be characterized by the effective Coulombic interaction within an implicit solvent.

% SI: get coexistence densities and dielectric constants. send solvation energy data to Sam. change LJ center to dipole center in P code, use all C.N. conditions for the P plot

\section{Conclusion}

In this work, we extended the shifted-dipole Stockmayer model to include ions in order to determine how dipolar asymmetry in the solvent impacts ion solvation. By varying the displacement of the solvent dipole within a minimal coarse-grained model, we distinguished the effects of solvent charge geometry from more chemically-specific interactions. As a result, we obtain a general understanding of how asymmetry in a polar solvent affects both the local structure around cations and anions and their corresponding solvation thermodynamics.

Structurally, the solvent with centered dipoles produces identical local environments around cations and anions. Shifting the dipole, however, breaks this symmetry and introduces a clear dependence on the sign of the charge. As the dipole shift $d^\ast$ increases, the solvent reorganizes so that anions are more closely coordinated by the positive dipole head, resulting in a compact first solvation shell, while cations are coordinated through the more inaccessible negative side of the solvent dipole, leading to a broader coordination shell. The corresponding trends in coordination number reinforce this asymmetry: with increasing dipole shift, the number of solvent particles around an anion decreases while that around a cation increases. Thus, the same molecular-scale displacement that breaks the head--tail symmetry of the solvent also changes how cations and anions are locally solvated. The consequences of this are first observed in the case of the polarizability and bound charge contribution from the dipoles, where, despite have a lower dielectric constant, shifted dipoles were found to be generally more effective at screening the local charge than the unshifted.

These findings are also reflected in the ionic solvation free energies. In the unshifted solvent, cation and anion ionic solvation free energies are identical, but increasing $d^\ast$ breaks this symmetry: anion solvation becomes progressively more favorable, while cation solvation becomes less so. This separation persists when the free energies are plotted with respect to the dipolar coupling parameter $\xi$, which accounts for changes in both dipole strength and density. The resulting trends thus point to the local asymmetry introduced by the shifted dipole, rather than density changes alone, as the source of the sign-dependent solvation response.

Overall, these results show that even a small molecular-scale displacement of the solvent dipole can produce a pronounced difference between positive- and negative-ion solvation, even when the ions themselves are otherwise identical. In the classical Born picture, ions of equal size and opposite charge would be expected to have the same solvation free energy because the response depends only on $q_i^2$ and the bulk dielectric constant. The asymmetry observed here therefore arises from molecular-scale solvent structure that is not captured by the simple continuum treatment. This then implies that the shifted solvent produces different effective Born solvation diameters for cations and anions despite their identical geometric diameters. More generally, these results demonstrate that sign-dependent solvation can emerge from solvent charge geometry alone, without requiring differences in ion size or chemically specific interactions.

Extending these findings, we next obtained the potentials of mean force between oppositely charged ions in a asymmetric dipolar solvent. At long-ranges, the cation--anion interaction did follow the expected behavior from the effective Coulombic two-body potential, where the lower dielectric constant of shifted dipolar solvents resulted in a stronger attraction. However, at short-ranges, the effects of the dipolar symmetry became more pronounced and reversed this trend.

These results help emphasize how implicit-solvent-type pictures of electrostatic interactions (such as the Born solvation free energy and the effective Coulombic potential) remain meaningful to understand the long-range behavior ions within dipolar solvents. However, even adding one small degree of asymmetry within the solvent results in effects which cannot be captured by such a simple picture. As such, future treatments of ions within asymmetric dipolar solvents will continue requiring the use of an explicit-solvent treatment\cite{blumAnalyticalSolutionMean1987,weiMeanSphericalApproximation1987}.

Nonetheless, as the shifted-dipole Stockmayer fluid isolates the sign-dependent effects within a minimal model, it provides a useful reference system for studies of ion-specific solvation, polar liquids, and electrolyte behavior. Future work could extend this framework to higher electrolyte concentrations, transport, and interfacial systems, where the same solvent asymmetry may influence both equilibrium structure and dynamics. Comparison with molecular solvents may help distinguish which aspects of sign-dependent solvation arise from solvent charge geometry and which require more detailed chemical interactions.

\begin{acknowledgments}
The authors would like to thank Dr. Benjamin Ye for helping develop many of the tools required for the analysis used in this work. S.V. was supported by the U.S. Department of Energy, Office of Science, Office of Advanced Scientific Computing Research, Department of Energy Computational Science Graduate Fellowship under Award Number DE-SC0022158 for part of this work. Partial support for this research is provided by Hong Kong Quantum AI Lab, AIR@InnoHK of the Hong Kong Government. B.Z. gratefully acknowledges support from the Harvey Mudd College start-up grant and NSF CAREER Award CHE-2337602.

\end{acknowledgments}

\section*{Supplementary Material}

The supplementary material contains details regarding the densities and dielectric constants for the pure shifted-dipole Stockmayer fluids considered in this work. We also provide details of an analytical theory for shifted-dipole polar fluids around ions.

\section*{Data Availability Statement}
The data that support the findings of this study are available from the corresponding author upon reasonable request.

\section*{References}
\bibliography{references_pierre}

\end{document}

\maketitle
\begin{center}
\small
\textsuperscript{1}Department of Chemistry, Harvey Mudd College, Claremont, CA 91711, United States\\[3pt]
\textsuperscript{2}Division of Chemistry and Chemical Engineering, California Institute of Technology, Pasadena, CA 91125, United States\\[3pt]
\textsuperscript{3}Department of Chemical and Biomolecular Engineering, Tandon School of Engineering, New York University, Brooklyn, NY 11201, United States\\[6pt]
\textsuperscript{\ensuremath{\dagger}}These authors contributed equally.\\
\textsuperscript{*}Corresponding authors: \href{mailto:bzhuang@g.hmc.edu}{bzhuang@g.hmc.edu}; \href{mailto:zgw@caltech.edu}{zgw@caltech.edu}
\end{center}

\section{Bulk Solvent Properties and Dielectric Response}
\label{sec:dielectric}

Simulation conditions and dielectric constants of the bulk shifted-dipole Stockmayer fluid for varying reduced dipole magnitude $\mu^*$ and dipole shift $d^*$ are reported in Table~\ref{tab:bulk_params}. All simulations contain $N=3000$ solvent particles at reduced temperature $T^*=1.0$ in cubic simulation boxes of side length $L^*$. The reduced number density is $\rho^*=N/(L^*)^3$. Dielectric constants $\varepsilon$ are reported with their corresponding statistical uncertainties.

\begin{table}[ht!]
    \centering
    \label{tab:bulk_params}
    \label{tab:bulk_dielectric}
    \begin{tabular}{ccccc}
        \hline\hline
        $\mu^*$ & $d^*$ & $L^*$ & $\rho^*$ &
        $\varepsilon$ \\
        \hline
        1.5 & 0.00 & 15.669 & 0.77976 & $39.50 \pm 1.05$ \\
            & 0.10 & 15.614 & 0.78813 & $28.33 \pm 0.65$ \\
            & 0.20 & 15.477 & 0.80920 & $18.10 \pm 0.35$ \\
            & 0.25 & 15.381 & 0.82440 & $14.69 \pm 0.29$ \\
        \hline
        1.8 & 0.00 & 15.262 & 0.84388 & $238.70 \pm 10.20$ \\
            & 0.10 & 15.220 & 0.85090 & $90.93 \pm 3.01$ \\
            & 0.20 & 15.079 & 0.87504 & $31.74 \pm 0.75$ \\
            & 0.25 & 14.987 & 0.89117 & $21.23 \pm 0.47$ \\
        \hline
        2.0 & 0.00 & 14.958 & 0.89642 & $11907.07 \pm 87.09$ \\
            & 0.10 & 14.987 & 0.89115 & $265.09 \pm 11.47$ \\
            & 0.20 & 14.853 & 0.91555 & $44.15 \pm 1.20$ \\
            & 0.25 & 14.763 & 0.93247 & $24.75 \pm 0.58$ \\
        \hline\hline
    \end{tabular}
\end{table}

\section{General theory of a shifted-dipole polar fluid}\label{sec:general}

Here, we derive the mean-field theory of a fluid whose molecular dipoles are shifted away from the centers of the particles. We first prescribe the density of molecular centers, so that we can describe the electrostatic response independently of the interactions that generate the density profile. We then consider a tagged ion and finally introduce the Lennard-Jones (LJ) interactions of the shifted-dipole Stockmayer fluid. Unlike prior treatments\cite{kusakaIoninducedNucleationDensity1995,langenbachCoOrientedFluidFunctional2017}, the displacement is retained explicitly throughout the derivation. Thus, we do not require the displacement to be small or the electric field to vary slowly over this distance.

\subsection{Molecular coordinates and microscopic densities}

A molecule is specified by the position of its center, $\vect{r}_i$, and the unit vector $\muhat_i$ pointing along its permanent dipole moment. The dipole rotates with the molecule and has fixed magnitude $\bar\mu$. Its moment and interaction-site position are:
\begin{align}
 \vect{\mu}_i &= \bar\mu\muhat_i,\\
 \vect{s}_i &= \vect{r}_i+d\muhat_i.\label{eq:site}
\end{align}
The scalar $d$ is the dipole displacement; positive $d$ shifts the site along the direction of the dipole. We use normalized orientational integration, with:
\begin{equation}
 \avint =1.
\end{equation}
We denote the microscopic number density of molecular centers by $\hat\rho$, and the microscopic polarization density by $\hat{\vect{p}}$:
\begin{align}
 \hat\rho(\vect{r}) &= \sum_{i=1}^{N}\delta(\vect{r}-\vect{r}_i),\\
 \hat{\vect{p}}(\vect{r})
 &=\sum_{i=1}^{N}\bar\mu\muhat_i
       \delta(\vect{r}-\vect{r}_i-d\muhat_i).
\end{align}
Here, $\delta$ denotes the three-dimensional Dirac delta function, and a hat distinguishes a microscopic density from its mean-field value. The prescribed external charge density is $\rho_c$. The total charge density, including the bound charge of the solvent, is:
\begin{equation}\label{eq:charge}
 \hat\rho_{\mathrm{tot}}(\vect{r})
 =\rho_c(\vect{r})-\nabla\cdot\hat{\vect{p}}(\vect{r}).
\end{equation}
In particular, $\rho$ is a number density, whereas $\rho_c$ and $\rho_{\mathrm{tot}}$ are charge densities. The distinction between the center and the dipole site is essential: the prescribed density refers to centers, while electrostatic interactions act at the dipole sites.

The Coulomb kernel and electrostatic energy are:
\begin{align}
 C(\vect{r}-\vect{r}')
 &=\frac{1}{4\pi\eps|\vect{r}-\vect{r}'|},\\
 \beta U_{\mathrm C}
 &=\frac{\beta}{2}\int\dd\vect{r}\int\dd\vect{r}'\,
 \hat\rho_{\mathrm{tot}}(\vect{r})C(\vect{r}-\vect{r}')
 \hat\rho_{\mathrm{tot}}(\vect{r}').\label{eq:coulomb}
\end{align}
Note that $\beta=1/(k_{\mathrm B}T)$, $T$ is temperature, $k_{\mathrm B}$ is Boltzmann's constant, and $\eps$ is the vacuum permittivity.

This charge representation is equivalent to the dipolar tensor representation used in our previous work\cite{varnerStockmayerFluidShifted2026}. In the absence of external charges, integration by parts gives:
\begin{align}
 \beta U_{\mathrm C}
 &=\frac12\int\dd\vect{r}\int\dd\vect{r}'\,
 \hat{\vect{p}}(\vect{r})\cdot\vect{T}(\vect{r}-\vect{r}')
 \cdot\hat{\vect{p}}(\vect{r}'),\\
 \vect{T}(\vect{r})&=-\beta\nabla\nabla C(\vect{r}),
\end{align}
however, we opt for the charge-based approach as it is more natural to include explicit charges through the charge density.

\subsection{Grand partition function and auxiliary fields}

We prescribe a center-density profile $\rho(\vect{r})$. Equivalently, the solvent volume fraction is $\phi_s=v\rho$, where $v$ is a reference molecular volume. For an incompressible solvent surrounding a cavity that represents a solute:
\begin{equation}\label{eq:incompressible}
 \phi_s(\vect{r})=1-\varphi_0(\vect{r}),
 \qquad \rho(\vect{r})=\frac{1-\varphi_0(\vect{r})}{v},
\end{equation}
where $\varphi_0$ is the imposed density profile of the solute. For an imposed liquid-vapor profile, $\rho$ can instead be specified directly. This constraint is imposed at the coarse-grained density level. We work in the grand canonical ensemble, with solvent chemical potential $\mu_{\mathrm{ch}}$; its subscript distinguishes it from the dipole moment. The constrained partition function is:
\begin{equation}\label{eq:Xi}
 \Xi=\sum_{N=0}^{\infty}\frac{e^{\beta\mu_{\mathrm{ch}}N}}{N!}
 \left(\prod_{i=1}^{N}\frac1v\int\dd\vect{r}_i
                   \int\frac{\dd\muhat_i}{4\pi}\right)\delta[\hat\rho-\rho]\,\exp[-\beta U_{\mathrm C}].
\end{equation}
The reference volume only fixes the convention for chemical potential. The density constraint can be resolved by a conjugate field $w$:
\begin{equation}
 \delta[\hat\rho-\rho]
 =\int\mathcal Dw\,
 \exp\left[i\int\dd\vect{r}\,w(\vect{r})
                   (\hat\rho(\vect{r})-\rho(\vect{r}))\right].
\end{equation}
We decouple the electrostatic energy through a Hubbard--Stratonovich transformation \cite{fredricksonEquilibriumTheoryInhomogeneous2005} with scalar field $\psi$:
\begin{equation}\label{eq:HS}
 e^{-\beta U_{\mathrm C}}\propto\int\mathcal D\psi\,
 \exp\left[-\frac{\beta\eps}{2}\int\dd\vect{r}\,|\nabla\psi|^2\right.\left.{}+i\beta\int\dd\vect{r}\,
                \hat\rho_{\mathrm{tot}}\psi\right].
\end{equation}
The use of proportionality omits a field-independent normalization. Without lack of generality, we can set the reference electrostatic potential to zero as $r\to\infty$ Using Eq.~\eqref{eq:charge}, the last integral becomes
\begin{align}
 \int\dd\vect{r}\,\hat\rho_{\mathrm{tot}}\psi
 &=\int\dd\vect{r}\,\rho_c\psi
   +\int\dd\vect{r}\,\hat{\vect{p}}\cdot\nabla\psi\\
 &=\int\dd\vect{r}\,\rho_c\psi
   +\sum_i\bar\mu\muhat_i\cdot\nabla\psi(\vect{r}_i+d\muhat_i).
\end{align}
All terms depending on the molecular coordinates and orientations can then be collected into the single-particle partition function:
\begin{align}
 Q[w,\psi]&=\frac1v\int\dd\vect{r}\,\avint
 e^{iw(\vect{r})+i\beta\bar\mu\muhat\cdot
                         \nabla\psi(\vect{r}+d\muhat)}\\
 &=\frac1v\int\dd\vect{r}\,e^{iw(\vect{r})}\avint
 e^{i\beta\bar\mu\muhat\cdot\nabla\psi(\vect{r}+d\muhat)}.\label{eq:Qaux}
\end{align}
Thus, $Q$ only depends on the field variables and is identical for all particles in the system, so the integrals over $N$ particles factorize simply into $Q^N$. Performing the sum grand-canonical sum over $N$ then gives:
\begin{equation}
 \sum_{N=0}^{\infty}\frac{(e^{\beta\mu_{\mathrm{ch}}}Q)^N}{N!}=\exp[e^{\beta\mu_{\mathrm{ch}}}Q].
\end{equation}
The grand-canonical partition function can now be expressed in terms of path integrals over the fluctuating field variables as:
\begin{equation}
      \Xi=\int\mathcal Dw\int\mathcal D\psi\,e^{-\beta H[w,\psi]},
\end{equation}
with the field-based Hamiltonian given by:
\begin{equation}
 \beta H[w,\psi]=\frac{\beta\eps}{2}\int\dd\vect{r}\,|\nabla\psi|^2
   -i\beta\int\dd\vect{r}\,\rho_c\psi+i\int\dd\vect{r}\,w\rho
   -e^{\beta\mu_{\mathrm{ch}}}Q[w,\psi].
\end{equation}

\subsection{Real saddle-point fields and the orientational integral}

At the saddle point, we introduce real fields through:
\begin{equation}
 iw=\omega=-\beta vu,\qquad \psi=i\phi.
\end{equation}
The field $u$ enforces the density constraint and has units of energy per volume; $\omega$ is dimensionless. The physical electrostatic potential and electric field are $\phi$ and $\vect{E}$, respectively, such that:
\begin{equation}\label{eq:efield}
 \vect{E}(\vect{r})=-\nabla\phi(\vect{r}).
\end{equation}
We use the standard bold uppercase symbols $\vect{E}$ for the electric field and $\vect{D}$ for the dielectric displacement field. Other vectors use lowercase bold symbols, including polarization $\vect{p}$.

We define the fugacity per volume as:
\begin{equation}
 \lambda=\frac{e^{\beta\mu_{\mathrm{ch}}}}{v}.
\end{equation}
Applying the saddle-point substitutions directly to the single-particle partition function gives:
\begin{align}
 Q[\omega,\phi]
 &=\frac1v\int\dd\vect{r}\,e^{\omega(\vect{r})}\avint
 e^{-\beta\bar\mu\muhat\cdot\nabla\phi(\vect{r}+d\muhat)}\\[5pt]
 &=\frac1v\int\dd\vect{r}\,e^{\omega(\vect{r})}\avint
 e^{\beta\bar\mu\muhat\cdot\vect{E}(\vect{r}+d\muhat)}.\label{eq:Qreal}
\end{align}
The Hamiltonian retains the same molecular term, now evaluated in the real fields:
\begin{equation}\label{eq:Hreal}
 \beta H[\omega,\phi]=\int\dd\vect{r}\,
 \left[-\frac{\beta\eps}{2}|\nabla\phi|^2
       +\beta\rho_c\phi+\omega\rho\right]-e^{\beta\mu_{\mathrm{ch}}}Q[\omega,\phi].
\end{equation}
The negative field-energy term follows from $\psi=i\phi$; the electrostatic field is a saddle-point variable. We find the mean-field solution by extremizing this Hamiltonian.

\subsection{Saddle-point equations}

Variation of $\beta H$ with respect to $\omega$ results in:
\begin{equation}
      \frac{\delta\beta H}{\delta\omega(\vect{r})}=\rho(\vect{r})-e^{\beta\mu_{\mathrm{ch}}}\frac{\delta Q}{\delta\omega(\vect{r})}=0
\end{equation}
where the functional derivative of $Q$ is given by:
\begin{equation}
 \frac{\delta Q}{\delta\omega(\vect{r})}=\frac1v e^{\omega(\vect{r})}\avint e^{-\beta\bar\mu\muhat\cdot\nabla\phi(\vect{r}+d\muhat)},
\end{equation}
ultimately yielding the following relationship for the density:
\begin{equation}
\rho(\vect{r})=\lambda e^{\omega(\vect{r})}\avint
 e^{\beta\bar\mu\muhat\cdot\vect{E}(\vect{r}+d\muhat)}.
 \label{eq:density-saddle}
\end{equation}
Thus, $\omega$ enforces the prescribed density through the full orientational integral. In any region that excludes solvent, $e^{\omega}$ goes to zero such that $\rho$ goes to zero.

Variation of $\beta H$ with respect to the electrostatic potential gives:
\begin{equation}
 \frac{\delta\beta H}{\delta\phi(\vect{s})}
 =\beta\eps\nabla^2\phi(\vect{s})+\beta\rho_c(\vect{s})
 -e^{\beta\mu_{\rm ch}}\frac{\delta Q}{\delta\phi(\vect{s})}=0.
 \label{eq:phi-variation-start}
\end{equation}
We next evaluate the molecular contribution $\delta Q/\delta\phi$ in this equation. To vary $Q$ with respect to $\phi$, let us switch from center coordinates to dipole-site coordinates at fixed orientation. The translation $\vect{s}=\vect{r}+d\muhat$ has unit Jacobian, so:
\begin{align}
 Q[\omega,\phi]
 &=\frac1v\int\dd\vect{r}\,\avint
 e^{\omega(\vect{r})-\beta\bar\mu\muhat\cdot\nabla\phi(\vect{r}+d\muhat)}\\[5pt]
 &=\frac1v\int\dd\vect{s}\,\avint
 e^{\omega(\vect{s}-d\muhat)-\beta\bar\mu\muhat\cdot\nabla\phi(\vect{s})}.
 \label{eq:translate}
\end{align}
At fixed $\omega$, differentiating this expression and integrating by parts yields:
\begin{align}
 \delta Q
 &=-\frac{\beta\bar\mu}{v}\int\dd\vect{s}\,
 \left[\avint\muhat\,
 e^{\omega(\vect{s}-d\muhat)-\beta\bar\mu\muhat\cdot\nabla\phi(\vect{s})}\right]
 \cdot\nabla\delta\phi(\vect{s})\\[5pt]
 &=\frac{\beta\bar\mu}{v}\int\dd\vect{s}\,
 \nabla\cdot\left[\avint\muhat\,
 e^{\omega(\vect{s}-d\muhat)-\beta\bar\mu\muhat\cdot\nabla\phi(\vect{s})}\right]
 \delta\phi(\vect{s}).
 \label{eq:Q-variation}
\end{align}
The surface term is taken to vanish. Comparing the coefficient of $\delta\phi(\vect{s})$ with the definition of the functional derivative gives:
\begin{equation}
 \frac{\delta Q}{\delta\phi(\vect{s})}
 =\frac{\beta\bar\mu}{v}\nabla\cdot
 \left[\avint\muhat\,
 e^{\omega(\vect{s}-d\muhat)-\beta\bar\mu\muhat\cdot\nabla\phi(\vect{s})}\right].
 \label{eq:Q-functional-derivative}
\end{equation}
Substituting Eq.~\eqref{eq:Q-functional-derivative} into Eq.~\eqref{eq:phi-variation-start}, and using $e^{\beta\mu_{\rm ch}}/v=\lambda$ and $\vect{E}=-\nabla\phi$, gives the full electrostatic functional derivative:
\begin{equation}
 \frac{\delta\beta H}{\delta\phi(\vect{s})}
 =\beta\eps\nabla^2\phi(\vect{s})+\beta\rho_c(\vect{s})
 -\beta\lambda\bar\mu\nabla\cdot\left[\avint\muhat\,
 e^{\omega(\vect{s}-d\muhat)+\beta\bar\mu\muhat\cdot\vect{E}(\vect{s})}\right]=0.
 \label{eq:phi-variation-explicit}
\end{equation}
We define the polarization $\vect{p}$ as the vector whose divergence gives the bound-charge contribution in Equation~\eqref{eq:phi-variation-explicit}, the Poisson equation:
\begin{equation}\label{eq:polarization}
 \vect{p}(\vect{s})=\lambda\bar\mu\avint\muhat\,
 e^{\omega(\vect{s}-d\muhat)+\beta\bar\mu\muhat\cdot\vect{E}(\vect{s})}.
\end{equation}
The shifted center factor counts molecules whose dipole sites lie at $\vect{s}$; the electric field acts locally at that site. The corresponding dipole site number density is given by:
\begin{equation}\label{eq:site-density}
 \rho_s(\vect{s})=\lambda\avint
 e^{\omega(\vect{s}-d\muhat)+\beta\bar\mu\muhat\cdot\vect{E}(\vect{s})}.
\end{equation}
The site density $\rho_s$ generally differs from the center density $\rho$, although their integrals over the full space count the same total number of molecules, since each molecule has one dipole.

Setting the electrostatic functional derivative to zero now gives:
\begin{align}
 \frac{\delta\beta H}{\delta\phi}
 &=\beta\eps\nabla^2\phi+\beta\rho_c-\beta\nabla\cdot\vect{p}=0,\\[5pt]
 -\eps\nabla^2\phi&=\rho_c-\nabla\cdot\vect{p}.\label{eq:Poisson}
\end{align}
Therefore, the complete self-consistent theory for a prescribed solute/solvent density profile is:
\begin{empheq}[box=\fbox]{align}
 \rho(\vect{r})
 &=\lambda e^{\omega(\vect{r})}\avint
 e^{\beta\bar\mu\muhat\cdot\vect{E}(\vect{r}+d\muhat)}
 =\phi_s(\vect{r})/v,\\[5pt]
 \vect{p}(\vect{r})
 &=\lambda\bar\mu\avint\muhat\,
 e^{\omega(\vect{r}-d\muhat)+\beta\bar\mu\muhat\cdot\vect{E}(\vect{r})},\\[5pt]
 \nabla\cdot\vect{D}&=\rho_c,
 \qquad\vect{D}=\eps\vect{E}+\vect{p},
 \qquad\vect{E}=-\nabla\phi.\label{eq:general-system}
\end{empheq}
These equations retain the displacement exactly and determine the pressure-like field $\omega$ and electrostatic potential $\phi$. The field is sampled at $\vect{r}+d\muhat$ in the center-density equation, whereas the center factor is sampled at $\vect{r}-d\muhat$ in the polarization equation. Both translations follow directly from $Q$.

For a prescribed center-density profile, the self-consistent equations can be solved by iteration, as follows. Start from an initial potential satisfying the electrostatic boundary conditions and calculate $\vect{E}=-\nabla\phi$. At the current field, the first equation in the boxed set determines $\omega(\vect{r})$ pointwise so that the prescribed center density is recovered. Use this field and $\omega$ in the second equation to calculate the polarization, including the translated center factor at $\vect{r}-d\muhat$. Then, solve the Poisson equation for an updated potential using this polarization and the specified charge density. Repeat these updates, mixing the new and previous fields if needed for stability, until the potential and polarization stop changing and the boxed equations are satisfied to the chosen tolerance.

\subsection{Relation to the small-displacement theory}

The small-displacement approximation can be obtained directly from $Q$. Expanding the field evaluated at the shifted site yields:
\begin{equation}
 \vect{E}(\vect{r}+d\muhat)
 =\vect{E}(\vect{r})+d(\muhat\cdot\nabla)\vect{E}(\vect{r})+O(d^2).
\end{equation}
For a planar field $\vect{E}=E_z(z)\hat{\vect{z}}$, let $t=\muhat\cdot\hat{\vect{z}}$ and let $\mathcal A$ be the interfacial area. The full molecular partition function and its local approximation are then:
\begin{align}
 Q[\omega,\phi]
 &=\frac{\mathcal A}{2v}\int\dd z\,e^{\omega(z)}
 \int_{-1}^{1}\dd t\,e^{\beta\bar\mu tE_z(z+dt)}\\[5pt]
 &\simeq\frac{\mathcal A}{2v}\int\dd z\,e^{\omega(z)}
 \int_{-1}^{1}\dd t\,
 \exp\left[\beta\bar\mu tE_z(z)
       +\beta\bar\mu d t^2\frac{\dd E_z}{\dd z}\right].
\end{align}
The second line recovers the quadratic orientational exponent of the earlier small-displacement treatment. In the present theory, we retain the translated field in the first line, so no gradient expansion or additional angular function is required.

\subsection{Mean-field grand free energy}

The mean-field grand free energy is the stationary value of the field Hamiltonian, $\Omega_{\mathrm{MF}}=H^*$. Integrating the density saddle-point equation over space yields:
\begin{equation}
 e^{\beta\mu_{\mathrm{ch}}}Q=\int\dd\vect{r}\,\rho(\vect{r}).
\end{equation}
Thus, the grand free energy on the density saddle point can be written directly as:
\begin{equation}\label{eq:Omega}
 \beta\Omega_{\mathrm{MF}}=\int\dd\vect{r}\,
 \left[-\frac{\beta\eps}{2}|\nabla\phi|^2+\beta\rho_c\phi
 +\rho(\omega-1)\right].
\end{equation}
The translational and orientational contributions are contained in $\omega(\vect{r})$. To eliminate this field, we solve the density saddle-point equation in Eq.~\eqref{eq:density-saddle} for $\omega$:
\begin{equation}
 \omega(\vect{r})
 =\ln\frac{\rho(\vect{r})}{\lambda}
 -\ln\left\{\avint
 \exp\left[\beta\bar\mu\muhat\cdot\vect{E}(\vect{r}+d\muhat)\right]\right\}.
 \label{eq:omega-from-density}
\end{equation}
Subtracting one and multiplying by $\rho(\vect{r})$ gives:
\begin{equation}
 \rho(\vect{r})[\omega(\vect{r})-1]
 =\rho(\vect{r})\left[\ln\frac{\rho(\vect{r})}{\lambda}-1\right]
 -\rho(\vect{r})\ln\left\{\avint
 \exp\left[\beta\bar\mu\muhat\cdot\vect{E}(\vect{r}+d\muhat)\right]\right\}.
 \label{eq:rho-omega-substitution}
\end{equation}
Substituting Eq.~\eqref{eq:rho-omega-substitution} into Eq.~\eqref{eq:Omega} yields the grand potential with $\omega$ eliminated:
\begin{multline}\label{eq:Omega-density}
 \beta\Omega_{\mathrm{MF}}=\int\dd\vect{r}\,
 \left[-\frac{\beta\eps}{2}|\nabla\phi|^2+\beta\rho_c\phi
 +\rho\left(\ln\frac{\rho}{\lambda}-1\right)\right]\\
 {}-\int\dd\vect{r}\,\rho(\vect{r})\ln\left\{\avint
 \exp\left[\beta\bar\mu\muhat\cdot\vect{E}(\vect{r}+d\muhat)\right]\right\}.
\end{multline}
The logarithm of the explicit angular integral is the single-molecule orientational contribution. Equation~\eqref{eq:Omega-density} is the grand potential after eliminating $\omega$. To obtain an alternative form on the electrostatic saddle point, we return to Eq.~\eqref{eq:Omega} and use Gauss' law, $\rho_c=\nabla\cdot\vect{D}$. Integration by parts gives:
\begin{align}
 \int\dd\vect{r}\,\rho_c\phi
 &=\int\dd\vect{r}\,(\nabla\cdot\vect{D})\phi
   =\int\dd\vect{r}\,\vect{D}\cdot\vect{E}.
 \label{eq:charge-integration-by-parts}
\end{align}
Here, the surface term $\int\dd\vect{S}\cdot\vect{D}\,\phi$ vanishes under the boundary conditions at infinity. Since $\vect{E}=-\nabla\phi$ and $\vect{D}=\eps\vect{E}+\vect{p}$, inserting Eq.~\eqref{eq:charge-integration-by-parts} into Eq.~\eqref{eq:Omega} gives
\begin{align}
 \beta\Omega_{\mathrm{MF}}
 &=\int\dd\vect{r}\left[-\frac{\beta\eps}{2}|\vect{E}|^2
 +\beta\vect{D}\cdot\vect{E}+\rho(\omega-1)\right]\\[5pt]
 &=\int\dd\vect{r}\left[-\frac{\beta\eps}{2}|\vect{E}|^2
 +\beta\eps|\vect{E}|^2+\beta\vect{p}\cdot\vect{E}
 +\rho(\omega-1)\right]\\[5pt]
 &=\int\dd\vect{r}\left[\frac{\beta\eps}{2}|\vect{E}|^2
 +\beta\vect{p}\cdot\vect{E}+\rho(\omega-1)\right].
 \label{eq:Omega-on-shell}
\end{align}
These are equivalent forms evaluated on the saddle-point fields: Eq.~\eqref{eq:Omega-density} has eliminated $\omega$, while Eq.~\eqref{eq:Omega-on-shell} retains it and uses the electrostatic saddle-point equation. For independent variations with respect to $\omega$ and $\phi$, we use the Hamiltonian in Eq.~\eqref{eq:Hreal} with the molecular term $Q$ retained.

\section{Point-dipole solvent around a tagged ion}\label{sec:ion}

\subsection{External charge, cavity, and radial electric field}

We now place a tagged ion at the origin. Its charge is $q_i=z_i e_0$, where $z_i$ is its signed valence and $e_0$ is the elementary charge. There are no other mobile ions. A point charge has:
\begin{equation}
 \rho_c(\vect{r})=q_i\delta(\vect{r}).
\end{equation}
The point-dipole solvent still requires a prescription that prevents dipole sites from reaching the singular charge. For example, we can exclude molecular centers from a spherical cavity of radius $R_c$:
\begin{equation}\label{eq:hard-cavity}
 \rho(r)=\rho_b\Theta(r-R_c),\qquad R_c>d,
\end{equation}
where $\Theta$ is the Heaviside function. Sites can then approach to $R_c-d$. Alternatively, the charge can be regularized. A convenient normalized Gaussian distribution of width parameter $b$ is:
\begin{equation}\label{eq:gaussian-charge}
 \rho_c(r)=\frac{q_i}{2^{3/2}b^3}
                 \exp\left(-\frac{\pi r^2}{2b^2}\right),
 \qquad\int\dd\vect{r}\,\rho_c=q_i.
\end{equation}
Charge regularization does not itself provide molecular excluded volume; the center-density profile remains an input in this section.

In spherical symmetry, the mean electric field, polarization, and displacement are radial:
\begin{equation}
 \vect{E}=E(r)\rhat,\qquad
 \vect{p}=p_r(r)\rhat,\qquad
 \vect{D}=D(r)\rhat.
\end{equation}
The scalar components are signed; in particular $E$ can be negative. Integrating Gauss' law over a sphere determines $D$:
\begin{align}
 \frac{1}{r^2}\frac{\dd}{\dd r}[r^2D(r)]&=\rho_c(r),\\
 D(r)&=\frac{1}{r^2}\int_0^r\dd r'\,r'^2\rho_c(r').\label{eq:Dr}
\end{align}
For a point charge, $D(r)=q_i/(4\pi r^2)$ for $r>0$. For Eq.~\eqref{eq:gaussian-charge},
\begin{equation}
 D(r)=\frac{q_i}{4\pi r^2}
 \left[\operatorname{erf}\left(\sqrt{\frac\pi2}\frac rb\right)
 -\sqrt{2}\frac rb e^{-\pi r^2/(2b^2)}\right].
\end{equation}

\subsection{Exact reduction of the shifted orientational integrals}

At a center position $\vect{r}=r\rhat$, let $t=\muhat\cdot\rhat$. Then, the radius of the shifted site and the projection of the dipole on the radial direction there are:
\begin{align}
 r_+(r,t)&=|\vect{r}+d\muhat|
          =\sqrt{r^2+2rdt+d^2},\\
 \muhat\cdot\widehat{\vect{r}+d\muhat}
 &=\frac{rt+d}{r_+(r,t)}.\label{eq:radial-projection}
\end{align}
Consequently, the field in the center weight is sampled and projected as:
\begin{equation}
 \muhat\cdot\vect{E}(\vect{r}+d\muhat)
 =E(r_+)\frac{rt+d}{r_+}.
\end{equation}
Conversely, for a specified site at radius $r$, the parent center lies at:
\begin{equation}
 r_-(r,t)=|\vect{r}-d\muhat|=\sqrt{r^2-2rdt+d^2}.
\end{equation}
In this reverse mapping, $t$ is the angle cosine relative to the radial direction at the site. It is the same integration variable in the angular quadrature, but it refers to a different observation point. The azimuthal integral is trivial in both cases, leaving $\frac12\int_{-1}^1\dd t$.

Substituting these geometric relations directly into $Q$ gives:
\begin{equation}\label{eq:Q-radial}
 Q[\omega,\phi]=\frac{4\pi}{v}\int_0^\infty\dd r\,r^2e^{\omega(r)}
 \frac12\int_{-1}^{1}\dd t\,
 \exp\left[\beta\bar\mu E(r_+)\frac{rt+d}{r_+}\right].
\end{equation}
The density and electrostatic variations therefore reduce to the following complete radial system:
\begin{empheq}[box=\fbox]{align}
 \rho(r)&=\frac{\lambda e^{\omega(r)}}{2}\int_{-1}^{1}\dd t\,
 \exp\left[\beta\bar\mu E(r_+)\frac{rt+d}{r_+}\right],\\[5pt]
 \rho_s(r)&=\frac{\lambda}{2}\int_{-1}^{1}\dd t\,
 e^{\omega(r_-)+\beta\bar\mu tE(r)},\\[5pt]
 p_r(r)&=\frac{\lambda}{2}\int_{-1}^{1}\dd t\,
 \bar\mu t
 e^{\omega(r_-)+\beta\bar\mu tE(r)}.\label{eq:radial-polarization}\\[5pt]
D(r)&=\eps E(r)+p_r(r).\label{eq:radial-system}
\end{empheq}
The prescribed $\rho(r)$ determines $\omega(r)$ through the first equation, while the remaining equations determine the field and the site response.
For a regular radial field, the expression at $r_+=0$ is interpreted through the vector limit. The point-charge cavity condition above avoids this singular sampling for occupied centers. The potential is obtained after solving for the field:
\begin{equation}
 \phi(r)=\int_r^\infty\dd r'\,E(r'),\qquad \phi(\infty)=0.
\end{equation}
Far from the ion, $\rho\to\rho_b$, $\omega\to\ln(\rho_b/\lambda)$, and:
\begin{equation}
 E(r)\sim\frac{q_i}{4\pi\eps\epsilon_{\mathrm r}^{\mathrm{MF}}r^2}.
\end{equation}
For a Gaussian charge and regular density, the radial field and polarization vanish at the origin. These inner and outer conditions complete the radial problem.

The angular distribution of centers at radius $r$ is a physical observable that can be evaluated after solving the field equations:
\begin{align}
 \rho(r,t)&=\lambda\exp\left[\omega(r)
 +\beta\bar\mu E(r_+)\frac{rt+d}{r_+}\right],\label{eq:angular-density}\\[5pt]
 \rho(r)&=\frac12\int_{-1}^{1}\dd t\,\rho(r,t).
\end{align}
This distribution describes orientations relative to the center-to-ion radial direction; it should be distinguished from the angular integral used to calculate polarization at a site. At $d=0$, reversing the ion charge and $t$ leaves the center distribution unchanged. At finite $d$, the radius $r_+$ and the projection in Eq.~\eqref{eq:radial-projection} change under $t\to-t$, allowing different structures around ions of equal size and opposite sign. A neutral cavity can also polarize the shifted-dipole solvent through its density gradient, so its reference state must be solved rather than assumed to have zero field.

\subsection{Electrostatic solvation and charging reference}
\label{sec:charging}

The vacuum field generated by the ion alone is:
\begin{equation}
 \vect{E}_{\mathrm{vac}}=\frac{\vect{D}}{\eps}.
\end{equation}
Given spherical symmetry, $D$ is fixed by the ion charge. Using $\vect{E}=(\vect{D}-\vect{p})/\eps$, the electrostatic terms in Eq.~\eqref{eq:Omega-on-shell} simplify as:
\begin{align}
 \frac{\eps}{2}|\vect{E}|^2+\vect{p}\cdot\vect{E}
 &=\frac{1}{2\eps}|\vect{D}-\vect{p}|^2
   +\frac1\eps\vect{p}\cdot(\vect{D}-\vect{p})\\
 &=\frac{|\vect{D}|^2}{2\eps}-\frac{|\vect{p}|^2}{2\eps}.
\end{align}
The first term is the vacuum self-energy. For a point charge, it diverges, so the solvent and vacuum expressions must use the same regularization before their difference is taken. After removing this term, the radial grand potential is:
\begin{equation}\label{eq:Omega-sub}
 \beta\Omega_{\mathrm{sub}}
 =4\pi\int_0^\infty\dd r\,r^2
 \left[-\frac{\beta p_r^2}{2\eps}+\rho(r)(\omega(r)-1)\right].
\end{equation}
The bulk volume contribution is subtracted when calculating an insertion free energy. The electrostatic charging free energy keeps the same cavity and reservoir and subtracts the neutral-ion state:
\begin{equation}\label{eq:charging-difference}
 \Delta\Omega_{\mathrm{chg}}(q_i)
 =\Omega_{\mathrm{sub}}(q_i)-\Omega_{\mathrm{sub}}(0).
\end{equation}
For fixed $\rho$, the translational terms cancel in this difference. For a compressible solvent they do not, because the solvent can reorganize upon charging.

\clearpage
\section{Results}\label{sec:results}
In the following sections, we analyze the results obtained for the shifted dipolar mean-field theory around central tagged neutral and charged particles, starting first with the polarization response of the solvent around a tagged neutral particle. 
\subsection{Radial polarization and screened charge}
\begin{figure}[!htbp]
\centering
\includegraphics[width=\textwidth]{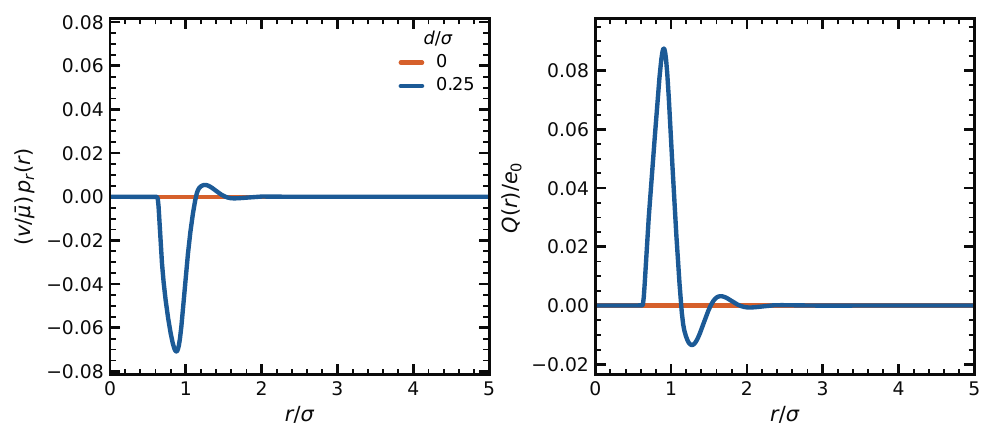}
\caption{Radial response of polar solvent around a neutral cavity. The panels show $(v/\bar\mu)p_r(r)$, the reduced radial polarization obtained from the polarization density in Eq.~\eqref{eq:radial-polarization}, and the running screened charge, which here is entirely due to induced bound charge. The orange curve is the centered-dipole case $d/\sigma=0$ and has no polarization; the blue curve is the shifted-dipole case $d/\sigma=0.25$ and shows the interfacial response generated by the prescribed cavity density profile. The enclosed charge necessarily returns to zero in the bulk.}
\label{fig:neutral-responses}
\end{figure}

We start by examining the polarization around a neutral particle to isolate the bare effect of interfacial polarization in this shifted-dipole model in the absence of any charge or external fields. In Figure~\ref{fig:neutral-responses}, we show the radial polarization profiles and running screened charge profile for a shifted and unshifted solvent around a neutral hard-wall cavity. In the case of the symmetric dipolar fluid (no shift), there is no polarization response at all, as expected by symmetry. On the other hand, the fluid with $d=0.25$ has a very significant polarization response, with the dipoles pointing toward the cavity. Interestingly, even at the mean-field level we do capture a small level of oscillation in the polarization. In Figure~\ref{fig:neutral-responses}b we plot the running screened charge, $Q(r)=4\pi\epsilon_0r^2E_r=Q_{\text{free}}-4\pi r^2 p_r(r)$.

\begin{figure}[!htbp]
\centering
\includegraphics[width=\textwidth]{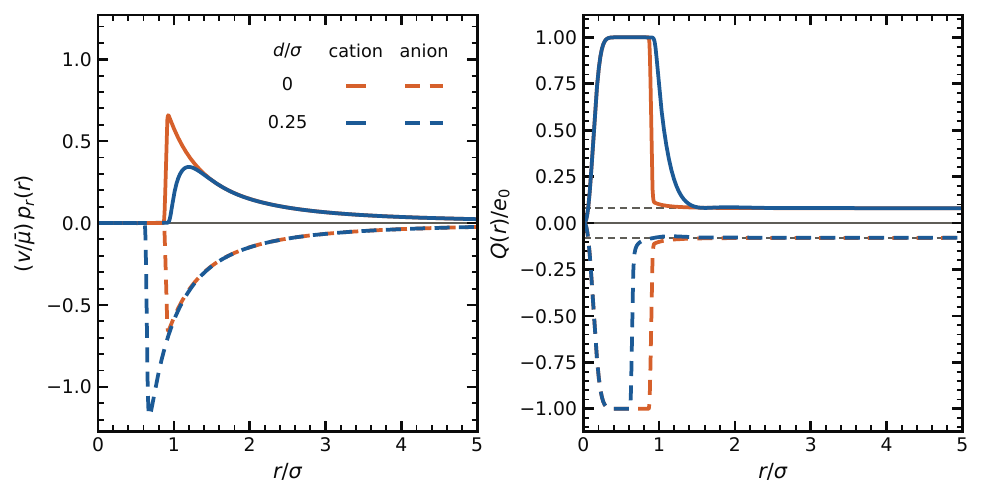}
\caption{Radial polarization and running screened charge around cations and anions in the prescribed-density point-particle theory. The polarization axis shows $(v/\bar\mu)p_r(r)$, the reduced radial component obtained from the polarization density in Eq.~\eqref{eq:radial-polarization}. The second panel shows $Q(r)=4\pi\epsilon_0r^2E_r$, the total enclosed free plus bound charge, in units of $e_0$. Line color distinguishes $d/\sigma=0$ and $0.25$; line style distinguishes cation and anion. Thin gray lines mark the charged-state asymptotes $Q/e_0=\pm1/\epsilon_{\mathrm{MF}}$.}
\label{fig:radial-responses}
\end{figure}

Next, we look at the polarization around a tagged cation and anion for both shifted and unshifted dipolar fluids. The polarization profiles and running screened charge are presented in Figure~\ref{fig:radial-responses}. In the case of no shift ($d=0$), the polarization around the cation is the exact mirror of the polarization around the anion due to symmetry. When a shift ($d=0.25$) is introduced, the polarization profiles become asymmetric. Namely, the polarization around the anion is stronger, with a peak shifted closer to the ion (smaller $r$) because the dipoles can place their heads closer to the source charge. Conversely, the cation induces a weaker polarization response that is further away from the charge source, since the dipole prefer to point away from the cation, and their tails will necessarily be further away from the cation in the shifted case. These effects further result in the charge being screened over a shorted distance for the anion than the cation in the shifted case, with the unshifted case lying in between. Note that in all cases the charge is not screened down to 0, but rather to $Q/e_0=\pm1/\epsilon_{\mathrm{MF}}$, as expected. Section~\ref{sec:dielectric} provides specific values of the dielectric constant for a given shift and dipole magnitude.

\clearpage

\subsection{Angular structure}
\begin{figure}[!htbp]
\centering
\includegraphics[width=\textwidth]{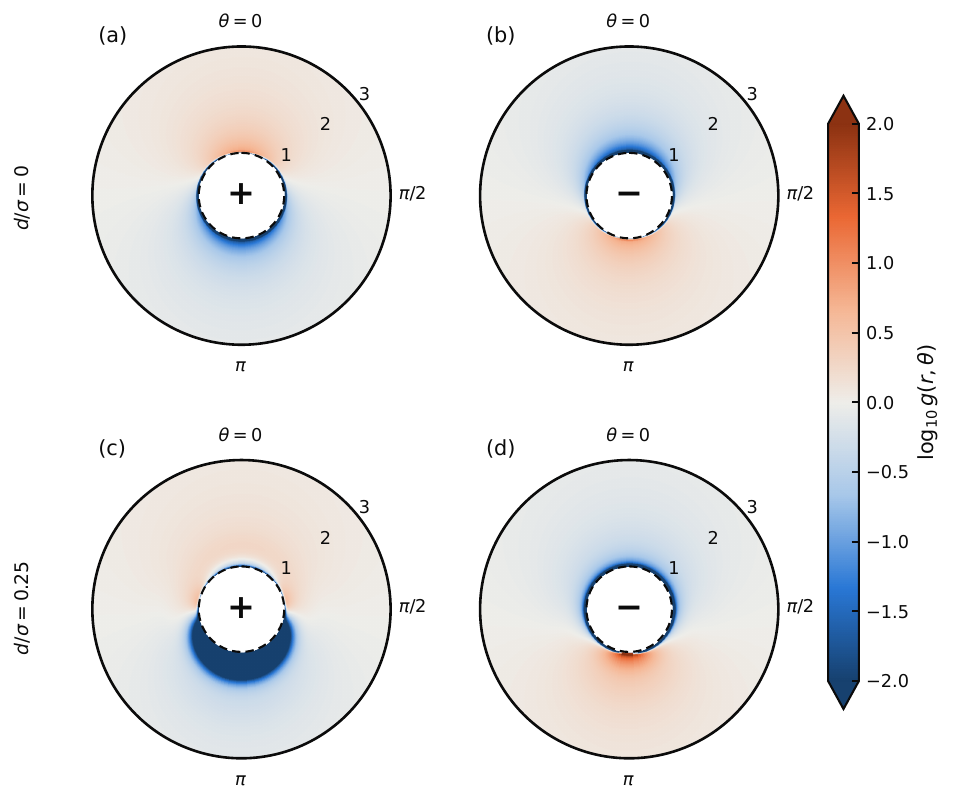}
\caption{Polar maps of the angular radial distribution function $g(r,\theta)=\rho(r,\theta)/\rho_b$ around cations (left column) and anions (right column), for $d/\sigma=0$ (top row) and $0.25$ (bottom row). The radial coordinate $r$ is the distance of the molecular center from the ion; the polar angle $\theta$ is measured from the outward radial direction, with $\theta=0$ at the top of each disk. Colors show $\log_{10}g$: blue indicates depletion and orange enrichment relative to the bulk value $g=1$, with the scale clipped at $\pm2$. The dashed circle marks the inner edge of the compact cavity profile used in Fig.~\ref{fig:radial-responses}.}
\label{fig:angular-rdf}
\end{figure}

For comparison to those results obtained in the main text, we obtain the polar maps of the angular radial distribution functions for solvent particles around a central ion given in Figure \ref{fig:angular-rdf}. These are in good agreement with the findings in the main text.
\clearpage
\subsection{Solvation free energy}
\begin{figure}[!htbp]
\centering
\includegraphics[width=0.78\textwidth]{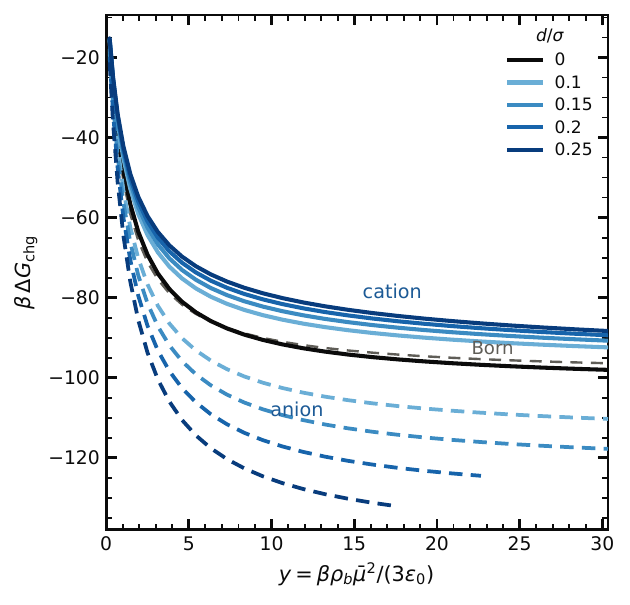}
\caption{Solvation free energy as a function of the bulk dipolar coupling $y=\beta\rho_b\bar\mu^2/(3\epsilon_0)$ for several dipole displacements. The dipole moment is varied at fixed temperature and molecular volume/density. Solid curves denote cations and dashed curves anions; the centered-dipole curve is common to both charge signs. The gray dashed curve is the Born estimate for the same solute size and charge.}
\label{fig:solvation-free-energy}
\end{figure}

By a similar token to the previous section, we compare the provide the predicted solvation free energies in dipolar solvents of varying dipole strength and shift in Figure \ref{fig:solvation-free-energy}. With the range of dipole strengths considered in the main text, there is good agreement in the magnitude of the solvation free energies.

\clearpage

\bibliography{references_pierre}
% \begin{thebibliography}{9}
% \bibitem{nakamura2012}
% I. Nakamura, A.-C. Shi, and Z.-G. Wang, ``Ion Solvation in Liquid Mixtures: Effects of Solvent Reorganization,'' \textit{Phys. Rev. Lett.} \textbf{109}, 257802 (2012), \href{https://doi.org/10.1103/PhysRevLett.109.257802}{doi:10.1103/PhysRevLett.109.257802}.
% \end{thebibliography}